\documentclass[usenatbib]{mn2e}
\usepackage{epsf,graphics,graphicx}
\usepackage{amsmath}
\usepackage{amssymb,latexsym,mathrsfs, bm}
\usepackage{color}
\usepackage{float}
\usepackage{natbib}
\usepackage{times}
\usepackage{color}

\def\eq#1{{Eq.~(\ref{#1})}}

\def\fig#1{{Fig.~\ref{#1}}}

\def\HI{{{\textrm{H}}~$\rm \scriptstyle I$}}
\def\HII{{{\textrm{H}}~$\rm \scriptstyle II$}} 
\def\HeI{{{\textrm{He}}~$\rm \scriptstyle I$}}
\def\HeII{{{\textrm{He}}~$\rm \scriptstyle II$}}
\def\HeIII{{{\textrm{He}}~$\rm \scriptstyle III$}}

\title[Quasar near-zones at $z \sim 6$]{Probing reionization using quasar near-zones at redshift $\mathbf{z \sim 6}$}
\author[Padmanabhan, Choudhury and Srianand]{Hamsa
Padmanabhan$^{1}$\thanks{Electronic address: hamsa@iucaa.ernet.in},
T. Roy Choudhury$^2$\thanks{Electronic address:
{tirth@ncra.tifr.res.in}},
 R. Srianand$^1$\thanks{Electronic address: {anand@iucaa.ernet.in}
}\\
$^{1}$ Inter-University Centre for Astronomy and Astrophysics, Pune 411007,
India\\
$^{2}$ National Centre for Radio Astrophysics, Tata Institute of Fundamental Research, Pune 411007, India}

\begin{document}
\date{ }
\maketitle

\begin{abstract}

{Using hydrodynamical simulations coupled to a radiative transfer code, we study the additional
heating effects in the intergalactic medium (IGM) produced by $z\sim6$ quasars in their near-zones. If helium is predominantly in \HeII\ to begin with, both normalization ($T_0$)
and slope ($\gamma$) of the IGM effective equation-of-state get
modified by the excess ionization from the quasars. 
Using the available constraints on $T_0$ at $z\sim6$, we discuss implications for the nature and epoch of \HI\ and \HeII\
reionization. We study the extent of the He~{\sc iii} region as a function of quasar age and
show, for a typical inferred age of $z\sim6$ quasars (i.e. $\sim 10^8$ yrs), it extends up to
80\% of the H~{\sc i} proximity region. For these long lifetimes, the heating effects can be detected even when all the  \HI\ lines from the proximity region are used. Using the flux and curvature probability
distribution functions (PDFs), we study the statistical detectability of heating effects
as a function of initial physical conditions in the IGM. For the present
sample size, cosmic variance dominates the flux PDF. The curvature statistics is more suited to
capturing the heating effects beyond the cosmic variance, even if the sample size is half of what is presently available.}
\end{abstract}



\section{Introduction}

Unravelling the process of reionization, which signals the end of the `dark ages' of our universe, is one of the current challenges of observational and theoretical cosmology. Two major milestones in the reionization history of the universe are those of hydrogen (\HI) and singly ionized helium (\HeII). Study of the evolution of hydrogen reionization combines observational evidences from various sources; optical probes include the Gunn-Peterson absorption troughs \citep{gunnpeterson} in the spectra of high-redshift bright sources such as (a) quasars \citep{fan, willott2007,mortlock2011}, (b) Lyman-$\alpha$ emitters \citep{kashikawa2006, stark2007, ouchi2010, nakamura2011} and (c) $\gamma$-ray bursts \citep[GRBs;][]{totani2006, kistler2009, ishida2011,robertsonellis2012}. The Thomson scattering optical depth measurements from the Cosmic Microwave Background (CMB) temperature and polarization power spectra are consistent with an instantaneous reionization at redshift $z \sim 11$ \citep{larson,planck, wmap7}, which may be interpreted as an estimate of the mean reionization redshift.  At radio frequencies, the redshifted 21-cm hyperfine line of neutral hydrogen promises a unique three-dimensional mapping of the epoch of reionization (EoR) of hydrogen \citep[for a review, see][] {furlanettorev}. All the available observations at present are consistent with an extended \HI\ reionization history that probably began at $z \sim 15$ and ended around $z \sim 6$ \citep{wyithe2003,trc2005, trc2006, pritchard2010, pritchard2010proc,mitra2011,mitra2012}.

The current observational probes of \HeII\ reionization include measuring the Gunn-Peterson absorption troughs in the \HeII\ Lyman-$\alpha$ forest \citep{jakobsen1994,zheng2004,reimers2005,shull2010, worseck2011,syphers2014}. These observations suggest that the EoR of \HeII\ is close to $z \sim 2.7$. The reionization of \HeII\ also leaves a thermal imprint on the hydrogen Lyman-$\alpha$ forest due to the additional heating effect on the velocity widths of the Lyman-$\alpha$ lines \citep{huignedin}.  The thermal evolution of the intergalactic medium (IGM) from $2 \leq z \leq 4.8$ has been probed using the observations of the Lyman-$\alpha$ forest \citep{ricotti2000a, schaye2000,mcdonald2001}. The velocity widths of the hydrogen Lyman-$\alpha$ forest lines seem to exhibit a sudden increase between redshifts $z \sim 3.5$  and 3, which may represent evidence for the reionization of \HeII. The inferred temperature measurements, taken in conjunction with the adiabatic cooling expected to occur after the reionization of hydrogen, also constrain the EoR of hydrogen to below $z \sim 9$ \citep{theuns2002}.
Recently, \citet{becker11} reported measurements of the IGM temperature from $2 \leq z \leq 4.8$ using the curvature statistic to quantify the temperature; their observations indicated gradual heating of the IGM from $z \sim 4.4$ towards lower redshifts, in contrast to the adiabatic cooling expected in single-step models of reionization. These measurements are consistent with an extended epoch of \HeII\ reionization starting probably at $z \gtrsim 4.4$ and terminating around $z \sim 3$.

Helium is expected to be singly ionized around the same time as the
hydrogen gets ionized, and first-generation galaxies are believed to be the likely sources for completion of hydrogen and \HeI\ reionization. In the single-step model of reionization, it is believed that massive, metal-free Population III stars \citep{ohetal2001,venkatesan2003} may have provided the hard photons required for \HeII\ reionization. In this model, a population of metal-free (Pop III) stars are required at redshifts $z > 6$ to reionize both \HI\ and \HeII. In the absence of a strong ionizing background for \HeII, it may recombine and hence to be reionized again at a lower redshift. Therefore, probes of intergalactic \HeII\ are important for understanding the role of Population III stars in the early reionization of \HeII\ and setting up a \HeII\ ionizing background prior to the quasar era (i.e. $z \sim 6$). Recently, there are indications of the presence of Population III stars even as late as $z \sim 3$ possibly due to inefficient transport of heavy elements and/or poor mixing that leave pockets of pristine gas even in chemically evolved galaxies \citep{jimenez2006, tornatore2007, inoue2011,cassata2013}. If, on the other hand, reionization took place as a two-step process (hydrogen first and \HeII\ later), quasars\footnote{{Strictly speaking, the term `quasar' is reserved for describing radio-loud quasi-stellar objects. However, as frequently done in the literature,  we will use the term `quasars' in this paper to indicate quasi-stellar objects, irrespective of their radio properties.}} are believed to be the most likely candidates for reionization of \HeII\ since their spectra are sufficiently hard. However, the number density of bright quasars peaks at $z \sim 2 - 3$ and decreases rapidly above $z \sim 4$ \citep{assef2011, masters2012}. Hence, in the two-step model of reionization, the final stages of \HeII\ reionization are expected to coincide with the peak of the quasar activity at  $z \sim 2 - 3$.

Quasar proximity zones\footnote{{Here, and in what follows, the term ``proximity zone'' or ``\HI\ proximity zone'' describes the region in the vicinity of the quasar where the ionizing flux from the quasar dominates the background flux.}}, where the excess ionization by the quasar allows the measurement of the velocity width of the Lyman-$\alpha$ line, have been used to probe the thermal state of the IGM at $z \sim 6$ \citep{bolton10}. This, in turn, can be used to probe the role of quasars in \HeII\ reionization. The IGM temperature in the near-zone\footnote{{Here, and in what follows, the term ``near zone'' refers to the region in the vicinity of the quasar within the \HeIII\ ionization front, where the heating effects are significant.}} is
influenced by both the existing background radiation as well as the additional radiation from the quasar itself. A first measurement of the near-zone
temperature around a quasar at redshift 6 has been reported in
\citet{bolton10} using Keck/HIRES data in combination with hydrodynamical
simulations. Recently, an additional source of heating has been observed in the
ionized near-zones of high-redshift quasars at  $z \sim 6$, which is attributed
\citep{bolton12} to the initial stages of helium
reionization around that redshift, since the excess heating can be easily accounted for if the \HeII\ is ionized by the quasar. The inferred excess temperature in the quasar near-zone can be used to place constraints on the epoch of \HI\ reionization \citep[see, for example,][]{ciardi,raskutti2012}.

In this paper, we explore several aspects of the additional heating effect in the near-zones of quasars at $z \sim 6$ using  the results of high-resolution hydrodynamical (SPH) simulations with {\sc gadget-2} \citep{gadget2}, and the ionization correction done using a 1D radiative transfer code which we have developed. The gas temperature in the general IGM is given by the assumed equation of state \citep{huignedin} and computed self-consistently for the near-zone of the quasar. We first validate our simulations by computing the additional temperature in the near zone for different initial equations of state of the general IGM, and different assumed values of the \HeII\ fraction prior to the active quasar phase. We obtain the expected relationship between the excess temperature and the initial \HeII\ fraction in the quasar near-zone, and  also find a connection between the magnitude of the steepening of the equation of state and the initial \HeII\ fraction. We then use our simulation results to measure the size of the region in the near-zone heated by the quasar in comparison to the \HI\ proximity zone, as a function of the age of the quasar. We also validate the usage of the flux and curvature statistics to measure the increased temperature in the near-zone of the quasar, and, in particular, address the effect of cosmic variance. For the flux statistics tests, we employ a number of pixels typical of the sample sizes in available observations of quasar near-zones at redshifts $\sim 6$. 
Using the Kolmogorov-Smirnov (KS) statistic to quantify the effect of the additional heating, and examining its variation with the parameters of the equation of state, $T_0$ and $\gamma$, we establish the connection between the thermal evolution of the IGM following the reionization of hydrogen, and the detectability of the additional heating in the quasar near-zone.  We also consider the possible dependence of the detectability of the additional heating effect on the assumed values of the background (metagalactic) photoionization rate of \HeII, which translates into varying the \HeII\ fraction in the near-zone of the quasar. This allows a connection to the effect of Population III stars on reionizing \HeII\ at redshifts $z > 6$ (which constrains the initial \HeII\ fraction in the quasar near-zone) in single-step reionization scenarios.

The paper is organized as follows: In Sec. \ref{sec:nummodel}, we describe our hydrodynamical simulations and the numerical
formalism for obtaining the simulated spectra in the quasar near-zone. In Sec. \ref{sec:validepen}, we provide a validation of our simulations by computing the excess temperature in the quasar near-zone for different values of the equation of state normalization, and the initial \HeII\ fraction, with comparison to the measured average temperature \citep{bolton12} in seven quasar near-zones at redshift $\sim 6$. We also describe the modification to the initial equation of state of the IGM due to the additional heating, and its dependence on the initial \HeII\ fraction in the quasar near-zone. In Sec. \ref{sec:results}, we describe the results obtained from our calculations as regards (a) the extent of the region around the quasar within which the additional heating is expected to contribute significantly, (b) the dependence of the additional heating effect in the near-zone on the initial equation of state of the IGM, quantified by the flux and curvature statistics, and (c) the dependence of the heating effect on the initial \HeII\ fraction in the near-zone, which is related to the single-step reionization by Population III stars. We then
summarize our findings in a brief concluding
section. Throughout this article, we assume the cosmological parameters $\Omega_m =
0.26$, $\Omega_{\Lambda} = 0.74$, $\Omega_b h^2 = 0.024$, $h = 0.72$, $\sigma_8
= 0.85$, and $n_s = 0.95$, which are consistent with the third-year WMAP and Lyman-$\alpha$ forest data \citep{seljak2006,viel2006}. The helium fraction by mass is taken to be 0.24
\citep{oliveskillman}.

\section{Brief description of numerical study}
\label{sec:nummodel}
\subsection{Hydrodynamical simulations and simulated spectra}

We perform cosmological hydrodynamical simulations using the parallel
smoothed-particle hydrodynamics (SPH) code {\sc gadget-2} \citep{gadget2}. We use two sets of simulations in this work: the lower resolution simulation contains $256^3$ each of gas and dark
matter particles in a periodic box of size $60 h^{-1}$ comoving Mpc, and the high resolution simulation contains $512^3$ each of gas and dark
matter particles in 
 a periodic box of size $10 h^{-1}$ comoving Mpc. In both cases, the gravitational softening length is 1/30th of the mean interparticle spacing, and initial
conditions are generated following the transfer function of Eisenstein and Hu
\citep{eisenstein}. Both sets of simulations are started at $z = 99$. Output baryonic density and velocity fields are generated at redshift $z \sim 6$.

Recently, it has been shown that when AGN feedback effects are taken into account in simulations, one finds that quasar host galaxies at redshifts $\sim 6$ are not `special' \citep{fanidakis2013}. It is now recognized that the existence of overdensities in the quasar near-zone can influence the background \HI\ photoionization rate measurements using the proximity effect \citep{rollinde2005, guimaraes2007, faucher2008}, but the thermal effects of choosing the quasar in a random position as compared to locating them in a high density environment may be minor \citep[][see Section 4.3 of the paper]{raskutti2012}. Observationally, \citet{willott2005} find no evidence of an overdensity of i-dropout galaxies around three $z \sim 6$ quasars, \citet{kim2009} find only two out of five quasar fields showing any evidence of overdensity, and \citet{banados2013}, studying the environment of a redshift 5.72 quasar, find no enhancement of Lyman-$\alpha$ emitters in the surroundings, compared to the blank fields. {For most part of this work}, we make the implicit assumption that quasars are not ``special'' and hence
do not arise preferentially in biased regions.  {However, we come back to this point and provide a qualitative discussion of the effects of locating the quasars in biased regions, in Sec. \ref{sec:othercont}.}

Lines of sight are extracted randomly in each simulation box at redshift
6, and the density and velocity fields along each line-of-sight is obtained. 
From the density grid of baryons in the simulation box, we compute the (physical) number densities of hydrogen and helium,
$n_{\rm{H}}$ and $n_{\rm{He}}$ (assuming the mass fraction $Y = 0.24$ of helium)
and then solve the equilibrium photoionization equations for \HI, \HeI\ and \HeII. Here, we explicitly assume
that the universe is already reionized and the IGM, assumed to be optically thin, is in
photoionization equilibrium with the background.  The background ionizing radiation is assumed to follow
the optically thin photoionization rates of
hydrogen and helium  as predicted by the  ``quasars + galaxies'' Haardt-Madau
background at redshift $\sim 6$, i.e. Table 3 of \citet{hm12}.
The value {of the background \HI\ photoionization rate} considered here is consistent at the  1$\sigma$ level with the results of the simulations of 
\citet{boltonhaehnelt07} and the observations of quasar {proximity zone sizes} in \cite{wyithe2011}. It is slightly higher than the value  ($1.57 \pm 0.62)  \times 10^{-13} {\rm{s}}^{-1}$, measured by \citet{calverley2011} using quasar proximity effects.  {The background \HeII\ photoionization rate, $\Gamma^{\rm bg}_{\rm HeII}$,} is known to have large fluctuations even at $z \sim 3$ due to the small number of ionizing sources within the characteristic mean free path of ionizing photons \citep[see, for example,][]{fardal1998, furlanetto2009, khaire2013}. At $z \sim 6$, this effect is expected to be severe, and the $\Gamma^{\rm bg}_{\rm HeII}$ we use is very small and should be treated as representative only. Later, we study the effect of varying this parameter on the results obtained.

In the absence of additional radiation from the quasars, we assign the gas temperature to each pixel by using the equation of state
of the photoionized IGM \citep{huignedin} with the normalization temperature $T_0 = 10^4$ K, and the slope $\gamma =
1.3$.
In principle,
$T_0$ and $\gamma$ at a given epoch can be fixed by comparing model predictions with
observations. Later, we also explore some models with physically motivated ranges in $T_0$ and $\gamma$ and draw conclusions regarding the epoch of \HI\ reionization.

We now evolve of temperatures and ion densities of hydrogen and helium (caused by ionization due to the quasar as well as the metagalactic background) along a line of sight with the quasar placed at the first gridpoint. The four
parameters, the temperature (obtained by using the equation of state) and the ion densities
of \HI, \HeII\ and \HeIII\ (obtained under the equilibrium conditions
with the photoionization rates from the background, i.e. without contribution from the quasar) are incorporated
as initial conditions.
The luminosity of the quasar at the Lyman edge,  $L_{\rm HI}$, is computed from the magnitude $M_{\rm AB} = -26.67$ at 1450 \AA\ (a typical magnitude for a luminous quasar at redshift $\sim 6$). We
assume the broken power law spectral index of $f_{\nu} \propto \nu^{-0.5}, 1050$ \AA \  $< \lambda
< 1450$ \AA, and $f_{\nu} \propto \nu^{-1.5}$ for $\lambda < 1050$ \AA. Hence, for the frequencies of interest, $f_{\nu} \propto \nu^{-\alpha_s}$ where $\alpha_s = 1.5$; the assumed spectral index is consistent with the inferred measurements \citep{wyithe2011} from observations of high-redshift quasar  {proximity zone} sizes. These parameters are then used
to derive the quasar contribution to the photoionization rates for \HI, \HeI\
and \HeII\ respectively. 

Since hydrogen is assumed to be highly ionized prior to the quasar being `switched on', the \HI\ ionization front from the quasar travels effectively at the speed of light. The region in the vicinity of the quasar in which the additional heating effects are expected to be significant may be characterized by the extent of the \HeIII\ region. To calculate the extent of this region, we track the location of the \HeII\ ionization front.
To do this, we use the relativistic equation of propagation of the ionization front modified to include the effects of optical depth:
\begin{equation}
 \frac{dR}{dt} = c\left(\frac{\dot{N}_{\rm eff} - 4 \pi R^3 n_{\rm He III} n_e \alpha_{\rm HeIII}/3}{\dot{N}_{\rm eff} + 4 \pi R^2 f_{\rm HeII} n_{\rm He} c - 4 \pi R^3 n_{\rm He III} n_e \alpha_{\rm HeIII}/3}\right)
 \label{front}
\end{equation} 
where $\dot{N}_{\rm eff} = \dot N e^{-\tau_{\rm HeII}}$ with $\dot{N}$ being the rate of production of \HeII-ionizing photons, and  $\tau$ being the optical depth at the \HeII\ edge at the distance $R$. The above equation is analogous to that used by \citet{icke1979} for the case of stellar Stromgren spheres, in which the optical depth effects are incorporated. Using the above equation, we can compute the time required by the \HeII\ front to reach a particular gridpoint under consideration. 
We can also compute the distance $R$ reached by the front after a time $t_Q$, where $t_Q$ is the lifetime of the quasar. This distance $R_{\rm He } = R(t = t_Q)$ is defined to be the location of the \HeII\ ionization front (or radius of the \HeIII\ ionized sphere) at the end of the quasar lifetime. We use this distance $R_{\rm He}$ to quantify the extent of the region in which additional heating effects are expected to be important, later in Sec. \ref{sec:rherh}.

{Our numerical procedure is described in detail in Appendix \ref{appendixa}. For the  evolution of the species densities and temperatures, we closely follow \citet{bolton07}. The radiative transfer implementation differs from \citet{bolton07} as regards the tracking of the ionization front. We have validated the front locations and speeds with Fig. 5 of \citet{mcquinn2012}, and the effect of the front propagation on the size of the near-zones is described in Sec. \ref{sec:rherh}.}

\subsection{Profile generation and statistics}

{We define the redshift grid along a line-of sight, using:
\begin{equation}
 x(z) = \int_0^z d_H (z')  dz'
 \label{redgrid}
\end{equation} 
where $d_H (z) = c(\dot{a}/a)^{-1}$ is the Hubble distance and $a$ is the scale
factor. 
}
Once we know the ion densities and gas temperatures at each pixel, following \citet{tirthanandtp1}, the Lyman$-{\alpha}$ optical depth due to
hydrogen at every redshift $z_0$ can be computed as:
\begin{eqnarray}
 \tau_{\rm \alpha}(z_0) &=& \frac{c I_{\alpha}}{\sqrt{\pi}} \int dx \frac{n_{\rm
HI}(x, z(x))}{b[x,z(x)] [1 + z(x)]} \nonumber \\
           && \quad \times \ V\left\{\alpha, \frac{c[z(x) -
z_0]}{b[x,z(x)](1+z_0)}  + \frac{v[x,z(x)]}{b[x,z(x)]} \right\}
\label{tau}
\end{eqnarray} 
where $b[x,z(x)] = \sqrt{2 k_B T[x,z(x)]/m_{\rm H}}$ is the thermal
$b$-parameter for hydrogen,  $V$ is the Voigt profile function, in which the
damping coefficient is $6.265 \times 10^8 {\rm{s}}^{-1}$, and $I_{\alpha} = 4.48 \times
10^{-18}$ cm$^{2}$ is related to the absorption cross-section
$\sigma_{\alpha}$ for the Lyman-$\alpha$ photons:
\begin{equation}
 \sigma_{\alpha} (\nu) = \frac{c I_{\alpha}}{b \sqrt{\pi}} V \left[\alpha,
\frac{c(\nu - \nu_{\alpha})}{b \nu_{\alpha}} \right]
\end{equation} 
where  $\nu_{\alpha}$ is the hydrogen Lyman-$\alpha$ frequency which corresponds
to the wavelength 1215.67 \AA. Using the above expression for the Lyman-$\alpha$ optical depth,  the simulated spectra are generated using $F = {\rm{exp}}(-{\tau_{\alpha}})$ for the flux $F$ at each pixel.\footnote{Though we do not convolve the spectra with instrumental broadening, this effect is expected to be negligible as compared to the thermal broadening effect which we are interested in.}
 We  mimic the noise by adding Gaussian
distributed noise having a signal-to-noise ratio {(SNR) 21}, equal to a typical SNR achieved for for $z \sim 6$ quasars with available instruments.
We generate spectra for a number of such lines of sight for the statistical analyses.
We consider two statistical indicators of the effect of the additional heating in this work : (a) the flux PDF statistics and (b) the curvature statistics. Note that \citet{bolton12} have used the cumulative distribution of velocity widths of Lyman-$\alpha$ lines obtained with Voigt profile fitting, to measure the temperature. However, unlike in the case of low redshift Lyman-$\alpha$ forest absorption, one will not be able to use higher Lyman-series lines to constrain the number of Voigt profile components. Hence, the derived $b$-distribution need not be well constrained. Therefore, in the present analysis, we explore the possibility of using the curvature statistics, that does not involve Voigt profile decomposition, to quantify the detectability of additional heating. Section \ref{sec:fluxstats} contains detailed descriptions of the flux and curvature statistics used to investigate the heating effect. 

\section{Excess heating in the quasar near-zones}
\label{sec:validepen}

In this section, we describe a validation of the numerical procedure by computing the additional heating effect and comparing it to the measured value of the average excess temperature in the near-zones of quasars in \citet{bolton12}. In particular, we investigate the effects of varying the normalization of the initial equation of state, and also the \HeII\ fraction in the vicinity of the quasar before the quasar is switched on. We explore how the combination of these parameters may be used to place possible constraints on the redshift of \HI\ reionization as well as single-step reionization models where \HeII\ is also ionized by massive stars. 

For this purpose, we employ the results of the $512^3$, $10 h^{-1}$ comoving Mpc box simulation with the quasar having a luminosity correponding to $M_{\rm AB} = -26.67$ at 1450 \AA, and a lifetime of 100 Myr. The initial equation of state parameters and the background photoionization rates are varied and the resulting final values of temperature as a function of $(1 + \delta)$, where, $\delta$ is the overdensity, are computed.

\subsection{Modifications to equation of state}

\begin{figure*}
 \begin{center}
   \includegraphics[scale=0.33, angle = 90]{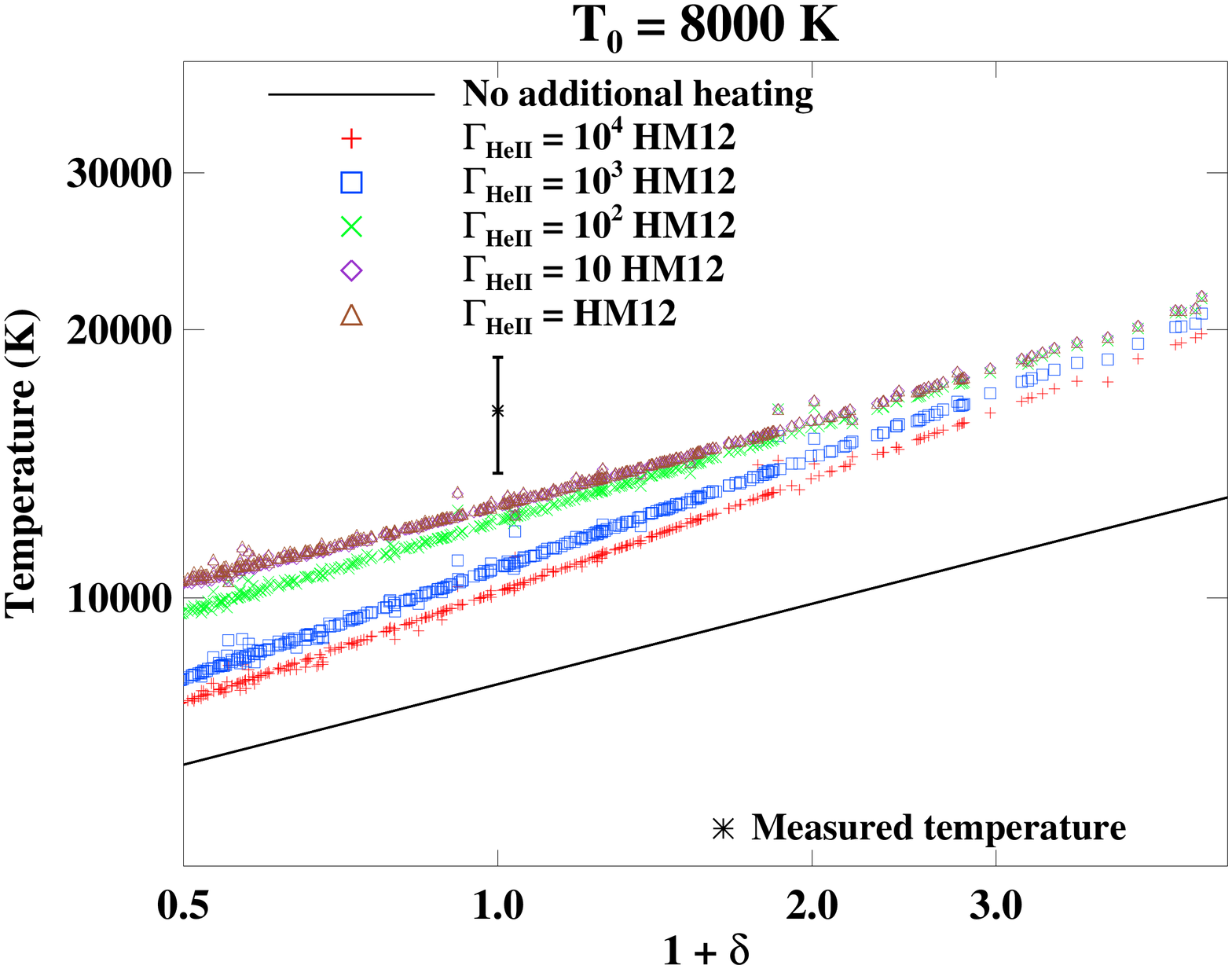} \includegraphics[scale=0.33, angle = 90]{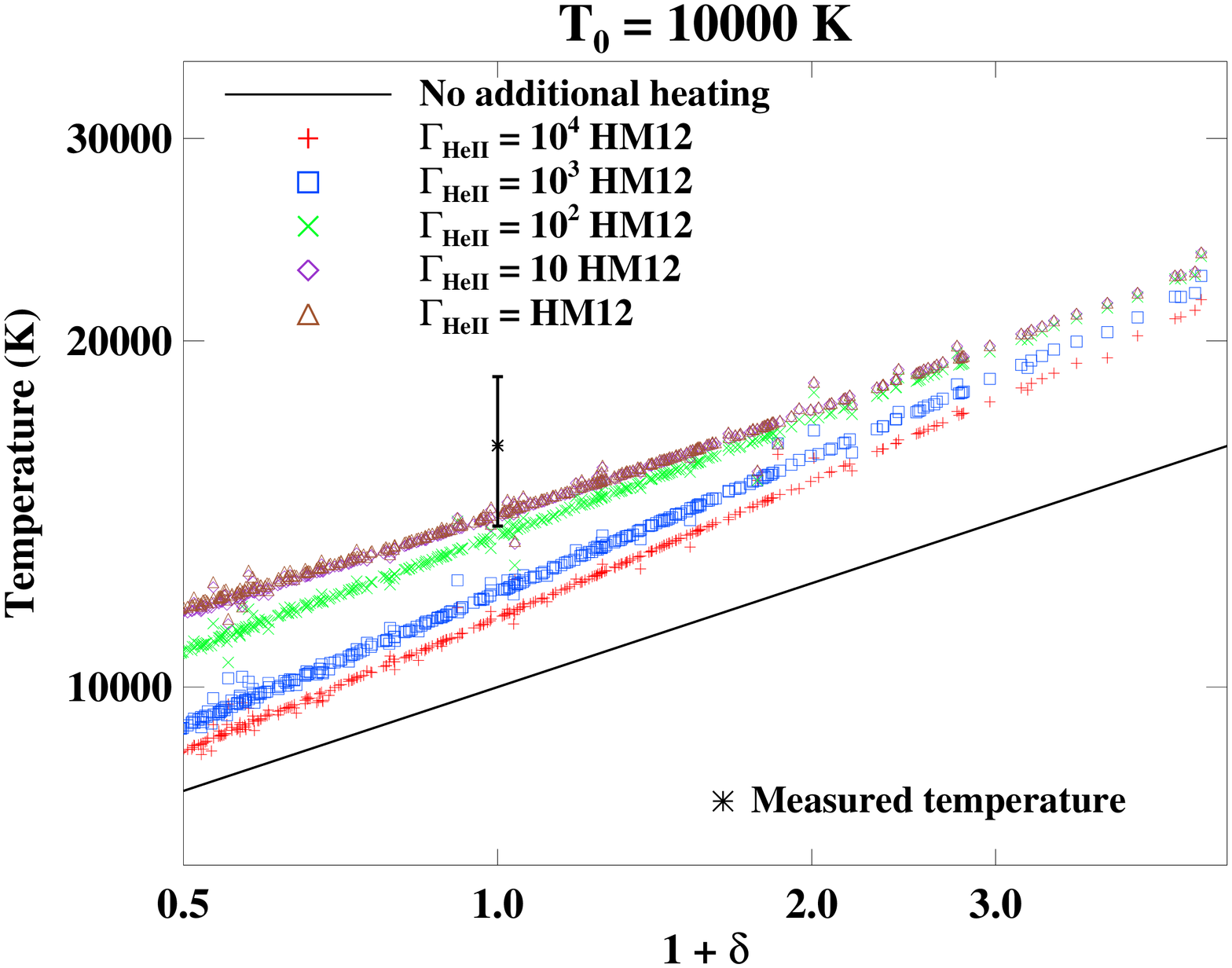} \vskip0.2in \includegraphics[scale=0.33, angle = 90]{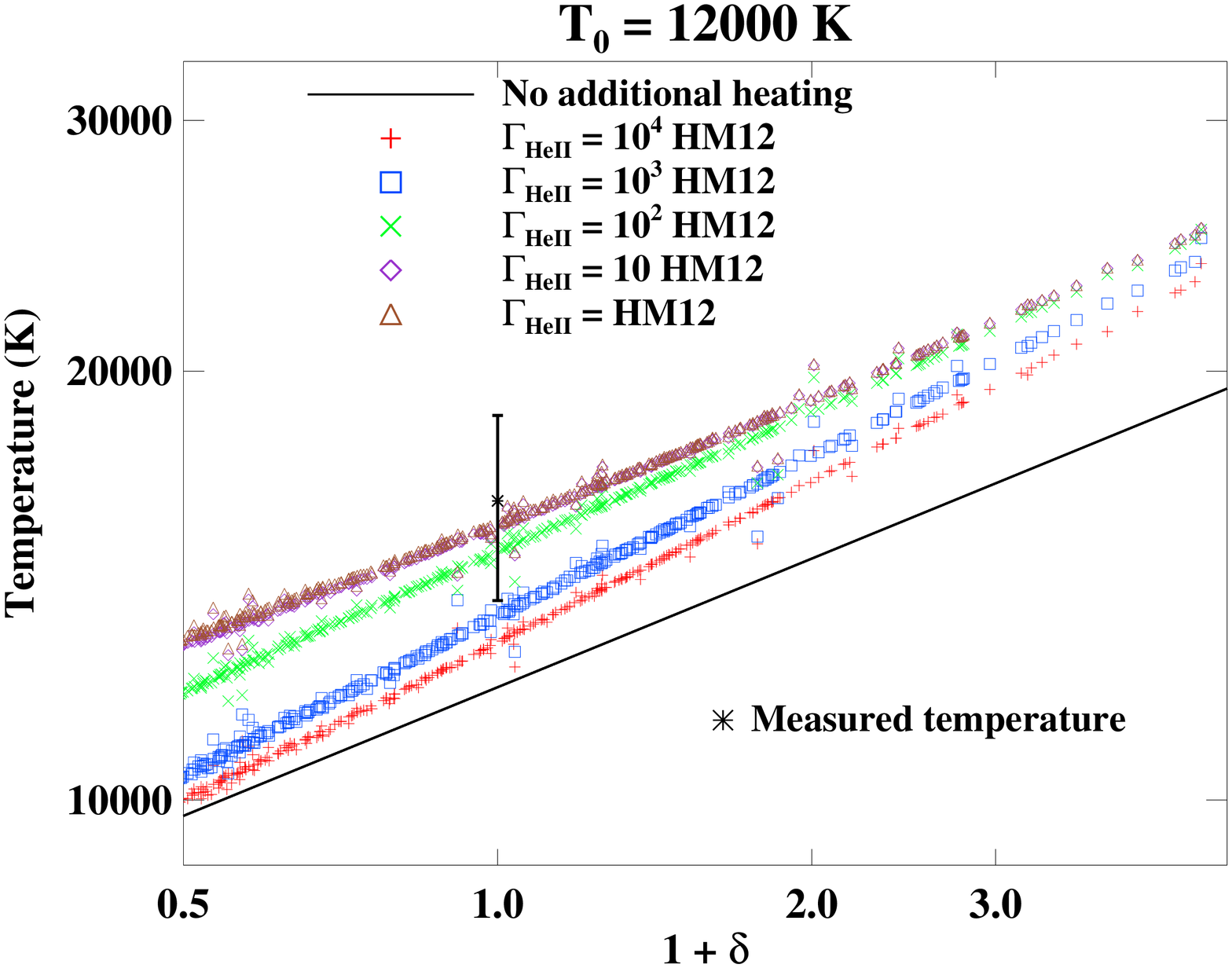} 
 \caption{The initial equation of state and the effect of the additional heating for different values of the background metagalactic photoionization rate, $\Gamma^{\rm bg}_{\rm HeII}$ (in s$^{-1}$). The normalization of the equation of state, $T_0$ is varied from 8000 - 12000 K. For each value of $T_0$, $\Gamma^{\rm bg}_{\rm HeII}$ is varied from $10^{4}$ HM12 to HM12, where HM12 is the Haardt-Madau background photoionization rate. This is equivalent to varying the initial \HeII\ fraction in the quasar vicinity from $x_{\rm HeII} \sim 0.05$ to $x_{\rm HeII} \sim 1$. In each figure, the asterisk with the error bar shows the measured average temperature in the near-zones of the seven redshift $\sim 6$ quasars considered in \citet{bolton12}.}
  \label{fig:constraints1}
 \end{center}
\end{figure*}

In Fig. \ref{fig:constraints1}, we have plotted the $T-(1+\delta)$ relation prior to and after additional heating by the quasar. We have chosen three different normalizations of the initial equation of state: $T_0 =$ 8000, 10000 and 12000 K, keeping the slope $\gamma = 1.3$ fixed. The range in $\delta$ plotted is from low to mildly nonlinear overdensities, and is representative of the range that contributes significantly to the intergalactic Lyman-$\alpha$ absorption seen in quasar spectra. For each value of $T_0$ considered, the parameter $\Gamma^{\rm bg}_{\rm HeII}$ is varied from $10^{4}$ HM12 to HM12, where HM12($\ =\ 4.42 \times 10^{-19}{\rm{s}^{-1}}$) is the value of the background \HeII\ photoionization rate computed by \citet{hm12}. This is equivalent to varying the initial \HeII\ fraction in the vicinity of the quasar from $x_{\rm HeII} \sim 0.05$ to $x_{\rm HeII} \sim 1$. We first describe the basic trends which are apparent in all the figures:

(a) For all values of $\Gamma^{\rm bg}_{\rm HeII}$ under consideration, there is an increase in the temperature. When $\Gamma^{\rm bg}_{\rm HeII}$ is higher (i.e. the initial $x_{\rm HeII}$ is close to 0.05), the temperature enhancement is less.  Also, irrespective of $\Gamma^{\rm bg}_{\rm HeII}$, the heated `equations of state' approach each other at high densities where the effects of recombination keep the \HeII\ fraction high, and hence the gas is heated to a higher temperature. Therefore, for higher $\Gamma^{\rm bg}_{\rm HeII}$, the measured value of $\gamma$ also becomes large (the ``heated'' equation of state acquires a steeper slope). 

(b) When $\Gamma^{\rm bg}_{\rm HeII}$ is very small (i.e. the initial $x_{\rm HeII}$ is close to 1), there is a uniform rise in temperature over the whole range of $\delta$ under consideration, i.e. we find a $\delta$-independent heating. This leads to the equation of state being shifted upward (i.e. only enhancement in $T_0$) with a negligible change in the slope. If indeed a major part of \HeII\ is ionized at $z \sim 6$ by the quasars, then our findings suggest that the \HI\ gas will still have some memory of the \HI\ reionization.

To summarize, there are two simultaneous trends which occur in the equation of state due to the decrease in $x_{\rm HeII}$: (a) a decrease in the normalization shift, and (b) an increase in the slope. We now consider these two trends separately, i.e. we explore the \textit{individual} change in the parameters $T_0$ and $\gamma$ ($\Delta T_0$ and $\Delta \gamma$) when the value of $x_{\rm HeII}$ is changed. 

\begin{figure}
 \begin{center}
   \includegraphics[scale=0.32, angle = 90]{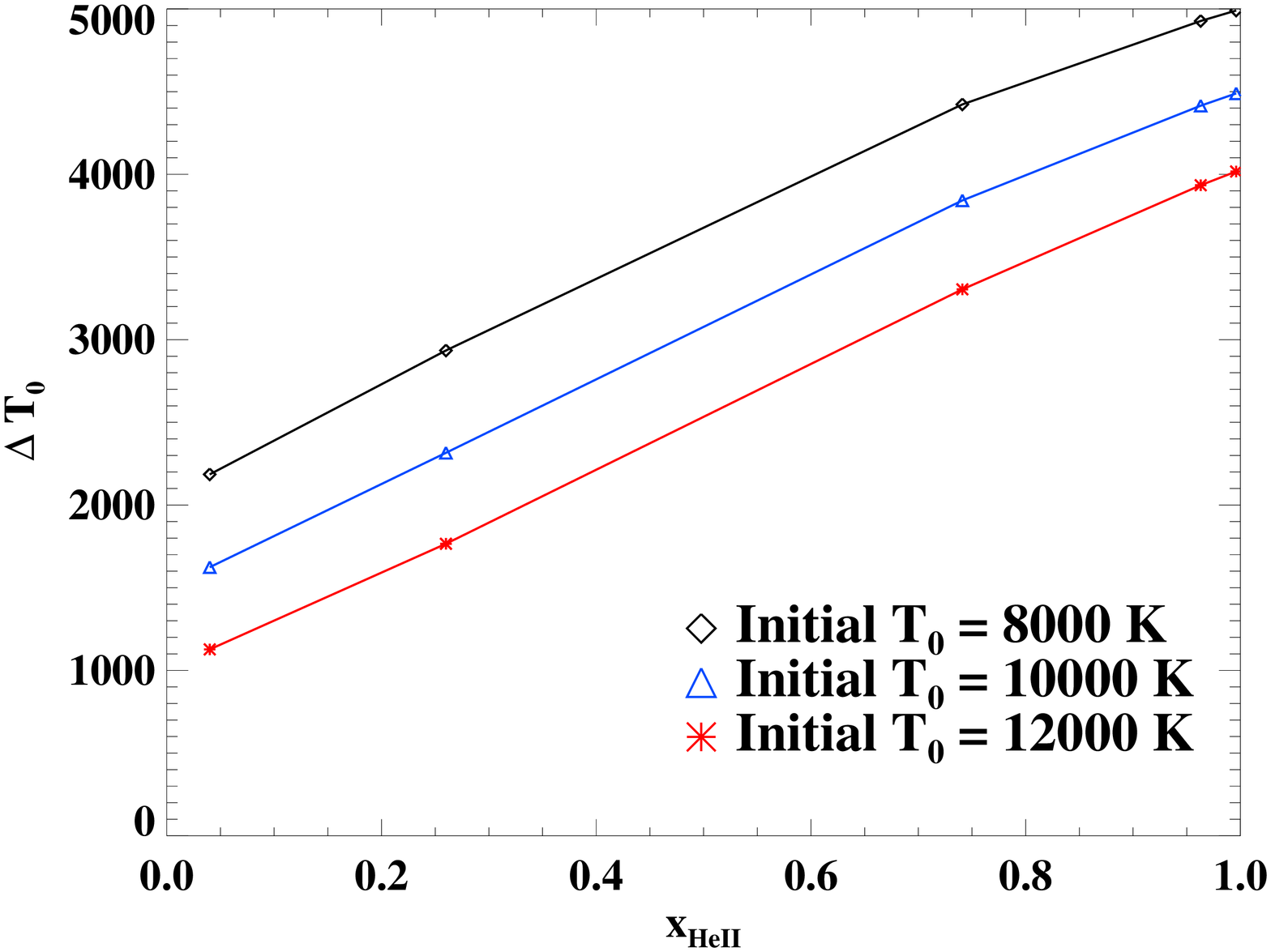} 
  \caption{The variation of $\Delta T_0$ with the initial $x_{\rm HeII}$. The relationship is linear, with the $\Delta T$ at a fixed $x_{\rm HeII}$ increasing with decrease in initial temperature.}
  \label{fig:changeinT0}
 \end{center}
\end{figure}

For each initial value of $T_0$ (8000 K, 10000 K and 12000 K), we plot the change in temperature at the mean density, $\Delta T_0$ against $x_{\rm HeII}$ for the five different values of $x_{\rm HeII}$ under consideration. This is shown in  Fig. \ref{fig:changeinT0}. It can be seen that  $\Delta T_0 \propto x_{\rm HeII}$ for all values of the initial $T_0$. This is in line with the analytic formulation provided in \citet{furlanetto2008} where it is argued that $\Delta T \propto x_{\rm HeII}$ (initial), where $\Delta T$ is the difference between the initial and heated temperatures. If we consider a fixed value of $x_{\rm HeII}$, for a higher initial $T_0$, the value of the $\Delta T_0$ is lower. This, again, is consistent with our previous findings that regions which are already `heated' can be additionally heated only to a limited extent.

\begin{figure}
 \begin{center}
   \includegraphics[scale=0.32, angle = 90]{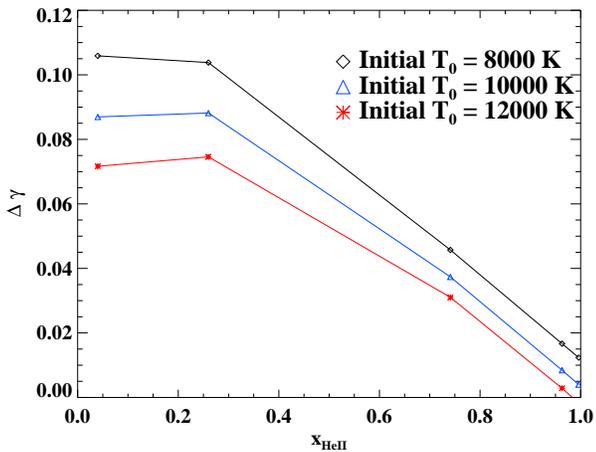} 
  \caption{The variation of $\Delta \gamma$ with the initial $x_{\rm HeII}$. The $\Delta \gamma$ at a fixed $x_{\rm HeII}$ increases with decrease in initial temperature, but reaches a maximum of about 0.1 at the lowest \HeII\ fractions under consideration.}
  \label{fig:changeingamma}
 \end{center}
\end{figure}

We now investigate the corresponding relationship for the case of the change in $\gamma$, i.e the $\Delta \gamma -x_{\rm HeII}$ relation. For this, we plot the difference $\Delta \gamma$ between the slopes of the `heated' and `initial' equations of state,  against $x_{\rm HeII}$, for the five different values of $x_{\rm HeII}$ under consideration. This is done for each initial value of $T_0$ (8000 K, 10000 K and 12000 K). The results are shown in  Fig. \ref{fig:changeingamma}. As expected, there is negligible change in $\gamma$ when the \HeII\ fraction is close to 1. We also note that for a fixed value of $x_{\rm HeII}$, the value of $\Delta \gamma$ is higher when the initial $T_0$ is lower. However, we see that the value of $\Delta \gamma$ reaches a maximum of about 0.1 at the lowest \HeII\ fraction and initial $T_0$ that we consider. The reason for this flattening is as follows: At high enough densities, all the curves in Fig. \ref{fig:constraints1} are constrained to follow the top curve due to recombination effects. At lower values of density, each curve in Fig. \ref{fig:constraints1} is shifted upward with respect to the initial equation of state, and the magnitude of this shift increases with increase in the value of $x_{\rm HeII}$. However, for low enough values of $x_{\rm HeII}$, both the `right top point' (which is constrained due to recombination effects) and the `left bottom point' (which is anchored close to the initial equation of state) are asymptotically fixed. This brings the slope to a near-saturation, which leads to the flattening out of $\Delta \gamma$. The maximum change in slope is greater if the shift in the overall normalization is higher, which happens if the initial $T_0$ is lower. Hence, the maximum value of $\Delta \gamma$ decreases with increase in the initial $T_0$, as we see in Fig. \ref{fig:changeingamma}. Our  $\Delta \gamma -x_{\rm HeII}$ relation above is analogous to the $\Delta T -x_{\rm HeII}$ noted in the literature. We infer that the value $\Delta \gamma \sim 0.1$ is representative of the maximum increase in the slope of the equation of state that may be achieved in physically feasible reionization scenarios. 

We speculate that the shifting upwards of the equation of state (which arises when the initial $x_{\rm HeII}$ values are high), may be easier to detect observationally than the (maximum) slope change of $\lesssim 0.1$ (which occurs when the initial $x_{\rm HeII}$ values are low). This also depends on how sensitive the statistical test used for distinguishability, is to the steepness of the equation of state, as compared to how sensitive it is to an overall increase in normalization. We will find, in the subsequent sections, that the curvature statistic is more sensitive to the expected shift $\Delta T_0 \sim 1000 - 5000$ K in the normalization of the equation of state, than to the expected change $\lesssim 0.1$ in its slope. 

\subsection{Implications of temperature measurements}

We now compare the results of our simulations with the available observations. At present, with a limited number of $z \sim 6$ quasars that are observed at high spectral resolution, constraints on the slope of the equation of state may be difficult. However, $T_0$ can be measured \citep[see][]{bolton12}. In what follows, we try to get constraints on the $\Gamma^{\rm bg}_{\rm HeII}$ using the available $T_0$ measurements. 
The measured average temperature (log $T$ (in K)$ \ =\ 4.21^{+0.06}_{-0.07}$) in quasar near-zones at redshift  $\sim 6$ \citep{bolton12} is indicated by the asterisk with error bar in each plot of Fig. \ref{fig:constraints1}. We note the following:

(a) If the initial equation of state has $T_0 = 8000$ K (a lower initial temperature), then the temperatures are lower than the $1\sigma$ lower bound on the measurement for all the $\Gamma^{\rm bg}_{\rm HeII}$ values under consideration. Thus it may be possible to rule out the corresponding reionization histories leading to this temperature prior to the switching on of the quasar. The temperature $T_0 = 8000$ K arises, for example, if we assume the instantaneous reionization followed by adiabatic cooling and compression, when the redshift of reionization of hydrogen is at $z_{\rm re} = 11$ with its associated temperature being $T_{\rm re} \sim 25000$ K.

(b) However, if the initial $T_0 = 10000$ K, then the $\Gamma^{\rm bg}_{\rm HeII}$ is constrained to $\lesssim 10^{-18}$ s$^{-1}$, which corresponds to $x_{\rm HeII} \gtrsim 0.96$, in order to be consistent with the measurements. The value of $T_0 = 10000$ K is, in turn consistent, with the reionization of \HI\ at $z_{\rm re} = 11$ and $T_{\rm re} \sim 30000$ K. These are physically acceptable redshifts and temperatures of \HI\ reionization.

(c) If the initial equation of state, on the other hand, has $T_0 = 12000$ K (a higher initial temperature), then the  $\Gamma^{\rm bg}_{\rm HeII}$ value is constrained to $\lesssim 10^{-16}$ s$^{-1}$ , which corresponds to $x_{\rm HeII} \gtrsim 0.26$, in order to be consistent with the measured temperature. The values of the initial $T_0 = 12000$ K and $\gamma = 1.3$ are difficult to reproduce with simple reionization models involving only adiabatic cooling and compression, but may arise in more complex models involving external sources of heating etc. In this case, the temperature measurement may be consistent with single-step models of reionization. It is to be noted that the additional heating effect is smaller for the case of higher initial $T_0$ than for the lower case. This leads to the curves being closer to each other in the bottom panel of Figure \ref{fig:constraints1}. In fact, this effect can be quantified using the curvature statistics by performing a Kolmogorov-Smirnov test between the `initial' and `heated' spectra, which we do and describe further in Section \ref{sec:fluxstats}. 

In this way, the exercise presented above validates our procedure and also captures the dependence of the heating to  (a) $T_0$, which connects up the heating effect to the epoch of hydrogen reionization in two-step models, and (b) $\Gamma^{\rm bg}_{\rm HeII}$, which connects to the possibility of single-step reionization of both \HI\ and \HeII. In any case, the prevalence of sufficiently hard sources at high redshifts substantially increases the $\Gamma^{\rm bg}_{\rm HeII}$ value and hence affects the temperature in the near-zone.  In the following sections, we quantify each of these effects, and also relate them to the \textit{detectability} of the additional heating using statistical analyses.

\section{Results}
\label{sec:results}

In the previous section, we have described in detail the modifications to the equation of state that occur due to the effect of the additional heating. We have also investigated the implications of the measured temperature in the near-zones of the quasars on the values of the various parameters of the IGM at that epoch. These point to constraints on both, the epoch of reionization of \HI\ as well as single-step models of reionization.
In the present section, we shall describe the main results of our simulations with respect to : (a) the relative extent of the He-heated region around quasars, compared to the \HI\ proximity zone, as a function of the age of the quasar, (b) the detectability of the additional heating effects as quantified by the flux and curvature statistics, and (c) implications for the detectability of additional heating in single-step reionization scenarios.

\subsection{Extent of additional heating around quasars}
\label{sec:rherh}

As the Lyman$-\alpha$ absorption from the general IGM at redshift 6 is optically thick, a profile analysis to estimate the gas temperature can be performed only in the quasar's proximity zone. In this zone, the \HI\ gas is highly ionized due to the excess ionization from the quasar. However, the fraction of this gas which is influenced by additional heat from the \HeII\ ionization by the quasar depends on where the \HeII\ front is located. This depends both on the quasar lifetime $t_Q$, as well as the line-of-sight optical depth for the \HeII\ ionizing photons. If the \HeIII\ front does not reach the edge of the \HI\ proximity zone for some reason, it would lead to dilution in the statistical tests to measure excess temperature. In order to provide estimates on the front location and the \HI\ proximity zone, a larger box-size (which includes these regions which are typically of the order of 8-9 proper Mpc) is required. Therefore, in this section, we address this issue using the lower resolution $256^3$, $60 h^{-1}$ comoving Mpc box simulation with the initial equation of state having $T_0 = 10^4$ K, $\gamma = 1.3$, and the quasar luminosity corresponding to $M_{\rm AB} = -26.67$ at 1450 \AA.

\begin{figure}
 \begin{center}
  \includegraphics[scale=0.4]{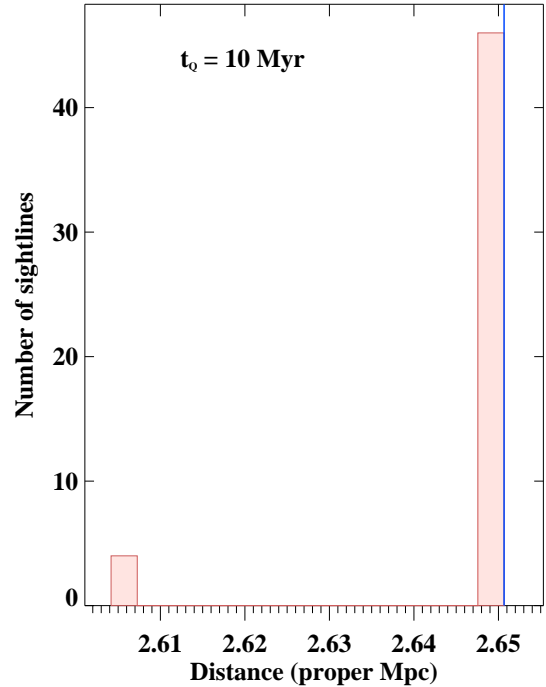} \includegraphics[scale=0.4]{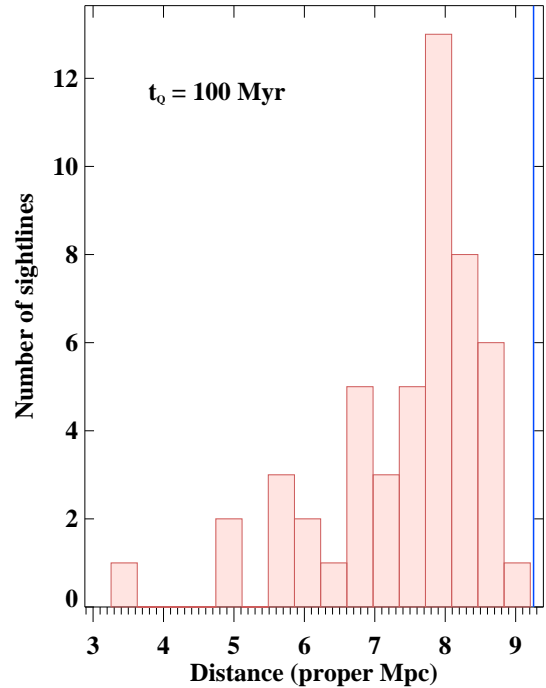}
  \caption{The extent of the \HeIII\ zone $R_{\rm He}$ for quasar lifetimes of  10 Myr (top panel), and 100 Myr (bottom panel). Each histogram comprises a total of 50 lines-of-sight. The blue vertical line shows the location of the \HeII\ front when the effect of optical depth is neglected, which represents the maximum extent of the \HeIII\ region for the given time.}
  \label{fig:RHe}
 \end{center}
\end{figure}

 Using \eq{front}, the equation of propagation of the \HeII\ ionization front that takes into account optical depth effects, we calculate the location $R_{\rm He} =R(t_Q) $ of the front  at the end of the quasar lifetime $t_Q$. 
The \HeII\ front location is computed for 50 random lines-of-sight extracted in the simulation box. We repeat the computation for two different values of $t_Q$, 10 Myr and 100 Myr\footnote{The assumed lifetimes of the quasar considered are indicative; at redshifts $z \sim 6$, measurements have placed the lifetimes of quasars at $\geq 10^7$ years \citep{haimancen2002, walter2003}.}, and the results are plotted in Fig. \ref{fig:RHe}. It can be seen that the extent of the \HeIII\ region (where additional heating of He, etc. are expected to be significant) increases as the quasar lifetime is increased, going up to about 8-8.5 proper Mpc from the quasar in a time interval of 100 Myr. The blue vertical line shows the maximum extent of the \HeIII\ region for a given $t_Q$ which occurs in the limit of zero optical depth. This is computed by setting $\tau_{\rm HeII} = 0$ in \eq{front}, so that $\dot N _{\rm eff} = \dot N$, where $\dot N$ is the rate of production of ionizing photons from the quasar. For quasar lifetimes of the order of 10 Myr, the optical depth effects are negligible and the mean location of the front is close to the maximum value that occurs in the limit of zero optical depth. For $t_Q \sim 100$ Myr, the front is able to travel a greater distance, but the optical depth effects begin to be important, and, on an average, the front reaches $\gtrsim$ 80\% of the maximum distance in about 66\% cases.

\begin{figure}
 \begin{center}
  \includegraphics[scale=0.4]{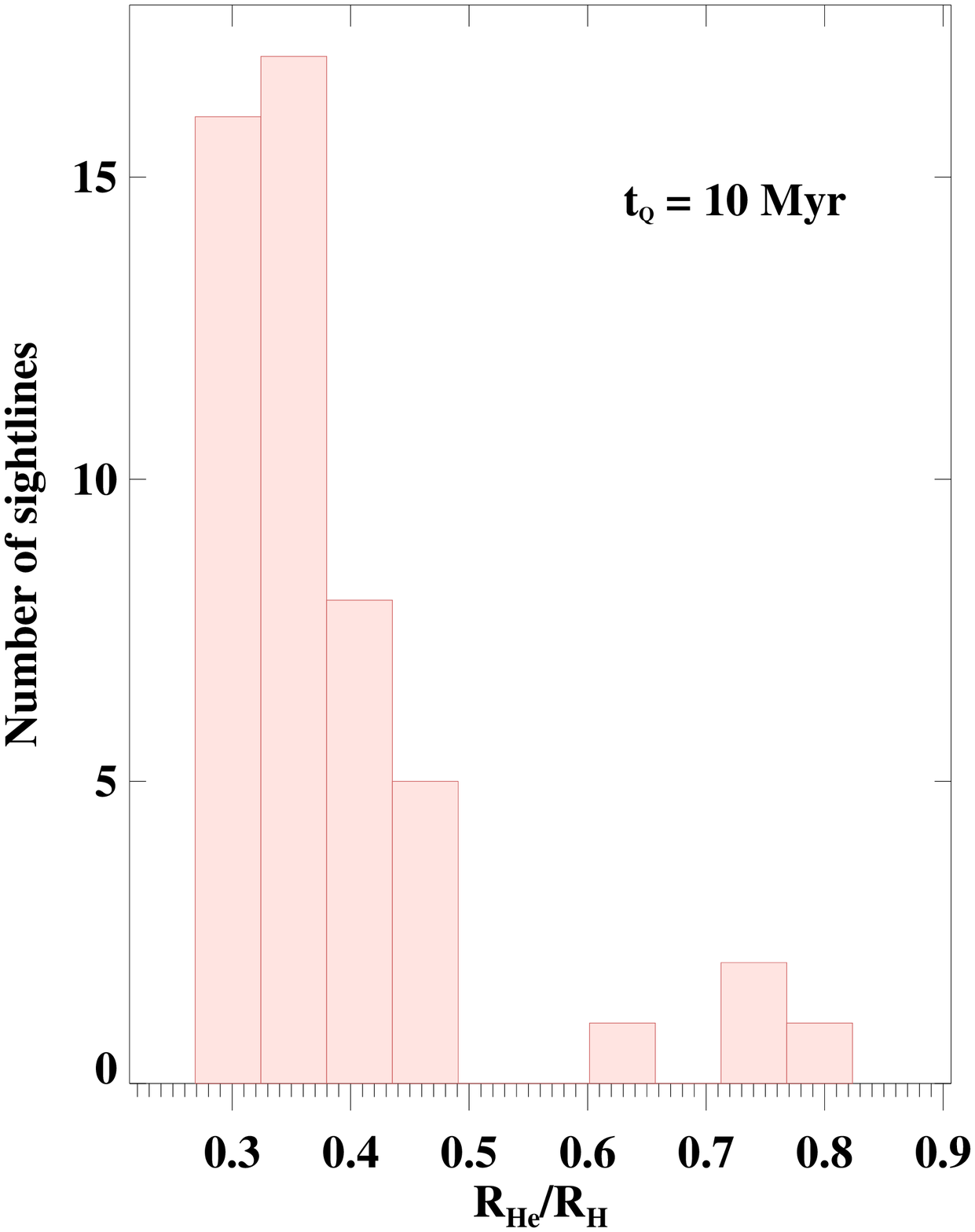} \includegraphics[scale=0.4]{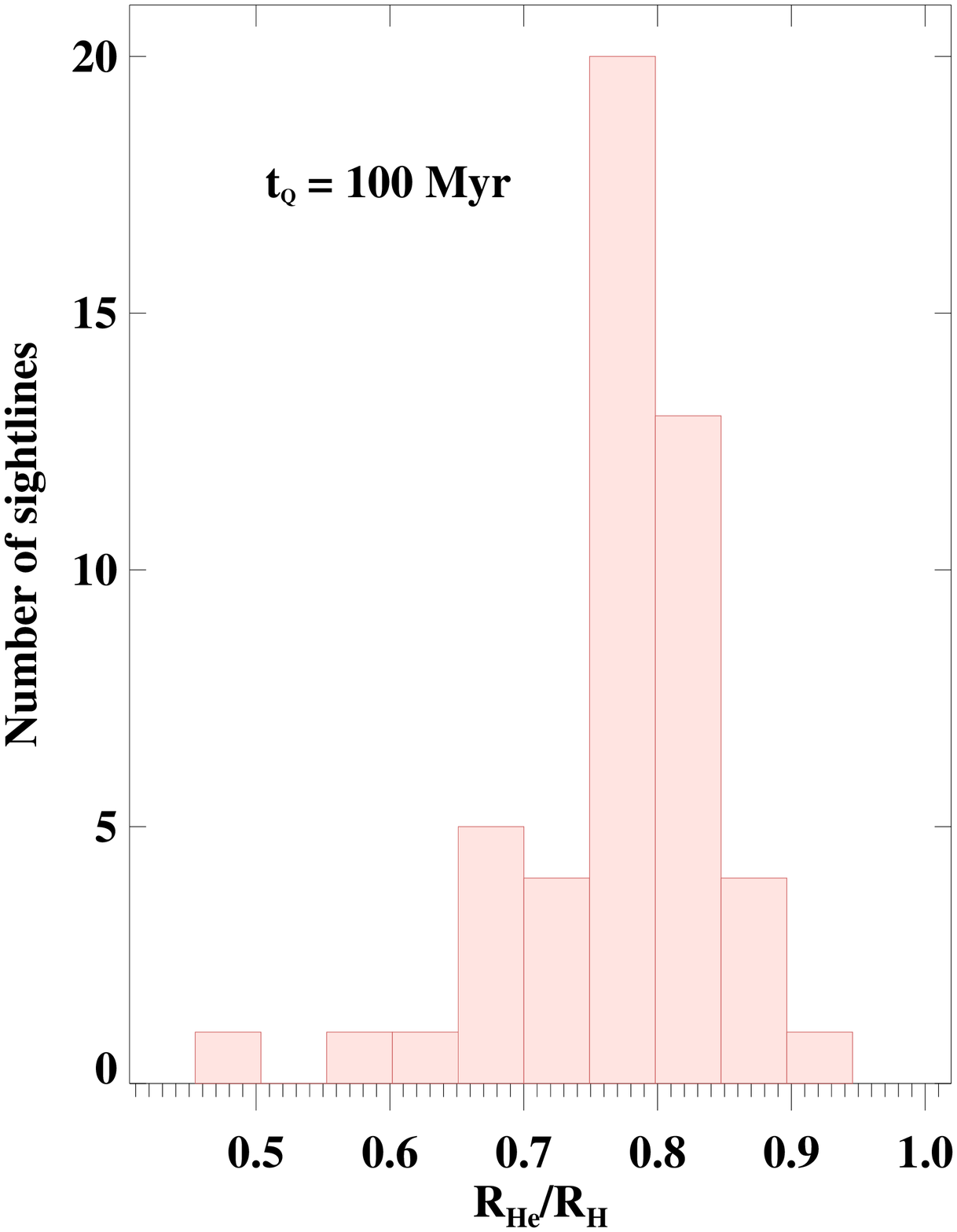}
  \caption{The \textit{relative} extent of the \HeIII\ zone with respect to the \HI\  proximity zone, $R_{\rm He}/R_{\rm H}$, for quasar lifetimes of  10 Myr (top panel), and 100 Myr (bottom panel). Each histogram comprises 50 lines-of-sight. As the quasar lifetime is increased, the relative extent of the \HeIII\ region also increases. For $t_Q \sim 10^8$ years (the typical inferred lifetime of the $z \sim 6$ quasar), more than 80\% of the \HI\ proximity zone is heated in 78\% of the sightlines.}
  \label{fig:RHebyRH}
 \end{center}
\end{figure}

We now consider the relative extent of the \HeIII\ region with respect to the \HI\ proximity zone of the quasar. Since one looks for the signatures of additional heating in the full \HI\ proximity zone of the quasar, it is important to quantify the extent of the region within this proximity zone in which additional heating effects due to ionization of \HeII\ are significant. The \HI\ proximity zone, $R_{\rm H}$, is defined through the relation $\Gamma_{\rm HI}^{QSO} (R_{\rm H}) = \Gamma^{\rm bg}_{\rm HI}$. The maximum value of $R_{\rm H}$ for the quasar luminosity under consideration and the background $\Gamma^{\rm bg}_{\rm HI}$, is $\sim 14$ proper Mpc from the quasar. The distance $R_{\rm He}$ is defined as $R_{\rm He} = R(t = t_Q)$ using \eq{front} with the optical depth effect taken into account. The ratio $R_{\rm He}/R_{\rm H}$, representing the relative extent of the \HeIII\ region within $R_{\rm H}$, is plotted as histograms 
in Fig. \ref{fig:RHebyRH} for the 50 lines-of-sight considered.  It can be seen that this ratio is about $30-35 \%$ for quasar lifetimes of the order of 10 Myr, but increases to about $80 \%$ for a quasar lifetime of $\sim 100$ Myr. This illustrates that the \HeII\ front covers about $80 \%$ of the \HI\ proximity zone of the quasar for $t_Q \sim 100$ Myr. 

The above result is closely connected with a related phenomenon of the  ``saturation'' or equilibrium value of the temperature in the region in which the heating effect is important. This saturation effect is seen as an increase in the temperature in a fairly distance-independent manner so that an equilibrium value is reached, after which there is little or no increase in the temperature over the timescales of interest for almost all gridpoints in the \HeIII\ region under consideration. This places a maximum bound on the temperature which the IGM may be heated to with ionization of both \HI\ and \HeII. This effect is reminiscent of the corresponding phenomenon in the interstellar medium where one finds the maximum temperatures to be $T_{\rm HI} \sim 20000$ K when \HI\ is ionized and $T_{\rm HeII} \sim 40000$ K when both \HI\ and \HeII\ are ionized; the exact values vary according to the detailed physics and optically thick/thin cases, but these numbers provide reasonable upper limits. In our present case the saturation is found to be achieved when the lifetime of the quasar is sufficiently high, $\sim 100$ Myr. Since the helium front covers about $80 \%$ of the \HI\ proximity zone within this time, the additional heating effect extends into a larger region and consequently, the rise in temperature is much more apparent, and fairly independent of distance. In contrast, for a quasar lifetime of 10 Myr, only about $30-35 \%$ of the \HI\ proximity zone near the quasar is influenced by the additional heating and it is possible that some of the pixels inside these regions have not yet reached the saturation in temperature. This means that for sufficiently long time scales ($\sim$ 100 Myr), the additional heating depends more on the initial IGM parameters and less on the distance from the quasar and the gas density. 
This turns out be important for the discussion in the following sections.

\subsection{Flux statistics and dependence on equation of state}
\label{sec:fluxstats}

In this section, we will explore some statistical tests to understand the sensitivity of the additional heating effect to the parameters of the general intergalactic medium at that epoch. For this purpose, we use the results of $512^3$ simulation box, which has a resolution of 2.65 km/s per pixel, and consider a quasar having a luminosity corresponding to $M_{\rm AB} = -26.67$ at 1450 \AA, and a  lifetime of 100 Myr. We consider two statistics which are both based on the observed hydrogen Lyman-$\alpha$ spectrum in order to quantify the additional heating effect, and the dependence on the equation of state parameters, $T_0$ and $\gamma$: (a) the probability distribution function (PDF) of the flux, and (b) the PDF of the flux curvature. We also consider the two-dimensional flux-curvature distribution. We  probe cosmic variance by using the same set of parameters, but different sets of lines-of-sight. 

The fiducial equation of state used for this purpose is $T_0 = 10^4$ K, $\gamma = 1.3$. The background $\Gamma^{\rm bg}_{\rm HI}$, $\Gamma^{\rm bg}_{\rm HeI}$ and $\Gamma^{\rm bg}_{\rm HeII}$ values correspond to those given by HM12 \citep{hm12} at redshift 6. The transmitted flux in the Lyman-$\alpha$ forest is sensitive to both, the temperature as well as the ionization state of hydrogen and therefore, to isolate the effect of additional heating around the quasar, we require the breaking of this degeneracy. For our chosen background photoionization rates, the spectrum when the quasar is not present is dark and hence featureless at redshift 6. Hence, it is impossible to compare the flux obtained from this spectrum with that when the quasar is present. Hence, we instead isolate the heating effect by generating a control sample (with the same initial conditions) of spectra with the temperature given by the initial equation of state and the ionization state being the same as that when the quasar is present. In other words, there is no He-related heating in the ``control'' sample.   Gaussian distributed noise is added to both the ``control" and the ``heated" spectra with a signal-to-noise ratio 21, mimicking the typical values in the observed HIRES quasar spectra. 

\begin{figure}
 \begin{center}
  \hskip-0.48in\includegraphics[scale=0.36, angle = 90]{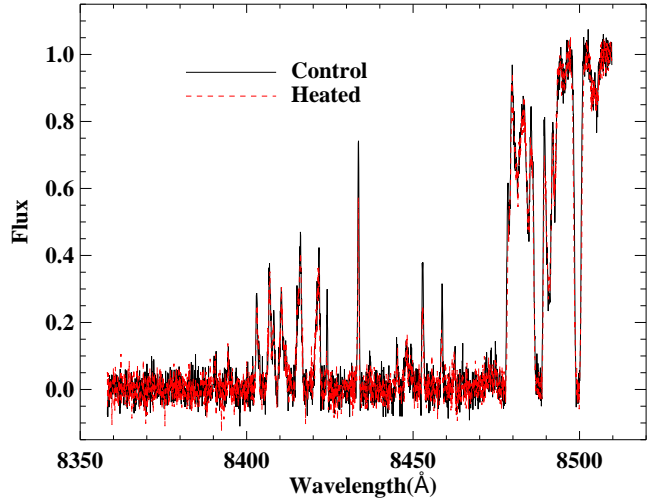}
  \caption{{Sample spectra, both heated (red dashed line) and control (black solid line) for a line-of-sight of having 2048 pixels drawn through the simulation box. The quasar lifetime is 100 Myr and the flux PDFs of the two samples are distinguishable with $94.5$ \%  confidence.}}
  \label{fig:longlos}
 \end{center}
\end{figure}

For all the statistical analyses, we replicate the typical sample size (total number of pixels) used  in the observational studies of the $z \sim 6$ quasars till now, since the spectral resolution in the observations is close to the resolution in our simulations. To take into account any distance-dependent effects, it may also be desirable to use a longer line-of-sight obtained by splicing together shorter sightlines available in the simulation box. However, we have seen in the previous section that for quasar lifetimes of the order of 100 Myr, the temperatures reach equilibrium and the heating effect becomes fairly independent of distance from the quasar. To illustrate this statistically, we implemented the numerical routine for the fiducial equation of state parameters, $T_0 = 10^4$ K and $\gamma = 1.3$ for a line-of-sight having length $40 h^{-1}$ comoving Mpc (constructed by splicing together four lines-of-sight of length $10 h^{-1}$ comoving Mpc each having 512 pixels), with the quasar lifetime of 100 Myr. The generated sample spectra, both heated (red) and control (black) are plotted in Fig. \ref{fig:longlos}. Five such  lines-of-sight were considered (so that the total sample size, $(2048 \times 5) \ {\rm{pixels}} \times 2.65$ km/s per pixel $\sim  7$ quasars $\times \ 3500$ km/s per quasar), and the flux PDF was generated for both the heated and the control spectra. The  flux PDFs for the control and the heated sample were compared using the Kolmogorov-Smirnov (KS) statistic, and they were found to be distinguishable with $94.5 \%$ confidence. 

This shows that the distinguishability of the samples is fairly independent of distance from the quasar if the quasar lifetime is of the order of 100 Myr. We also noticed that the temperature enhancement is fairly independent of the distance of the pixel from the quasar, for this case. On the other hand, if the same exercise is repeated for a quasar lifetime of 10 Myr, it is found that the sample with additional heating resembles the control sample very strongly and the two flux PDF distributions are distinguishable only at the $15 \%$ level. This is to be expected since, as we have seen in Sec. \ref{sec:rherh}, the helium front travels to only about $30-35\%$ of the hydrogen near-zone in this lifetime and hence the additional heating effect is confined to a small part of the line-of-sight under consideration. \footnote{{It is assumed that the quasar shines with constant flux during the entire lifetime for the purpose of the simulations. For long timescales ($\gtrsim$ few Myr), the quasar light curves cannot be constrained using direct observations. However, these and subsequent results depend upon the integrated thermal effects throughout the active lifetime of the quasar. The luminosity of the quasar used in the simulations is to be taken an estimate of the average luminosity of the quasar throughout its active lifetime.}}

We thus infer that for sufficiently long timescales of $\sim$ 100 Myr, the actual location of the pixel with respect to the quasar may not be as relevant as other parameters such as the initial equation of state as far as the heating effect is concerned. For this reason, in all the further statistical studies, we will use 20 lines-of-sight of length $10 h^{-1}$ Mpc each comprising 512 pixels, which replicates the sample size in the observations of the quasar spectra \footnote{We are not concerned with the spatial density correlations in the present study.}. 

\subsubsection{Flux PDF statistics}
\label{sec:fluxpdf}

{We compare two samples of 20 lines-of-sight each having 512 pixels\footnote{Due to the limited box size of the simulation, about 20 pixels at the extreme of the box have slight errors in the Lyman-$\alpha$ optical depth due to the incompleteness of the integral in the Voigt profile generation. For the statistical tests, therefore, we discard these 20 pixels (equivalent to about 50 km/s) at the extreme of the box.} for the ``control'' and ``heated'' spectra generated, using the Kolmogorov-Smirnov (KS) statistic. Note that apart from the additional heating, all other parameters of the heated model are identical to the ``control'' one. The results for the cumulative flux distributions are plotted below in Fig. \ref{fig:eosfluxpdf1} and Fig. \ref{fig:eosfluxpdf2}, along with the KS statistics `$d$' (the maximum separation between the two cumulative probability distributions) and `prob' (the probability that the two samples come from the \textit{same} parent distribution) in each case. In Fig. \ref{fig:eosfluxpdf1}, the temperature at mean density is fixed at $T_0 = 10^4$ K and $\gamma$ is increased from 1.1 to 1.5. It can be seen, that the distributions for the samples with and without additional heating may be distinguished with $\sim 100\%$ confidence when $\gamma = 1.1$, but only with  $69.78\%$ confidence when $\gamma = 1.5$. Hence, a higher slope of the initial equation of state leads to a greater resemblance to the control sample. In Fig. \ref{fig:eosfluxpdf2}, the slope is fixed at $\gamma=1.3$ and $T_0$ is varied from 8000 K to 12000 K. The flux PDFs for the sample with and without additional heating are distinguishable at the $99.79\%$ level when $T_0 =$ 8000 K, but only at the $87.13 \%$ level when $T_0 =$ 12000 K. Hence, if the initial $T_0$ is larger, the distinguishability of the two samples becomes poorer. }

\begin{figure}
 \begin{center}
  \includegraphics[scale=0.35, angle = 90]{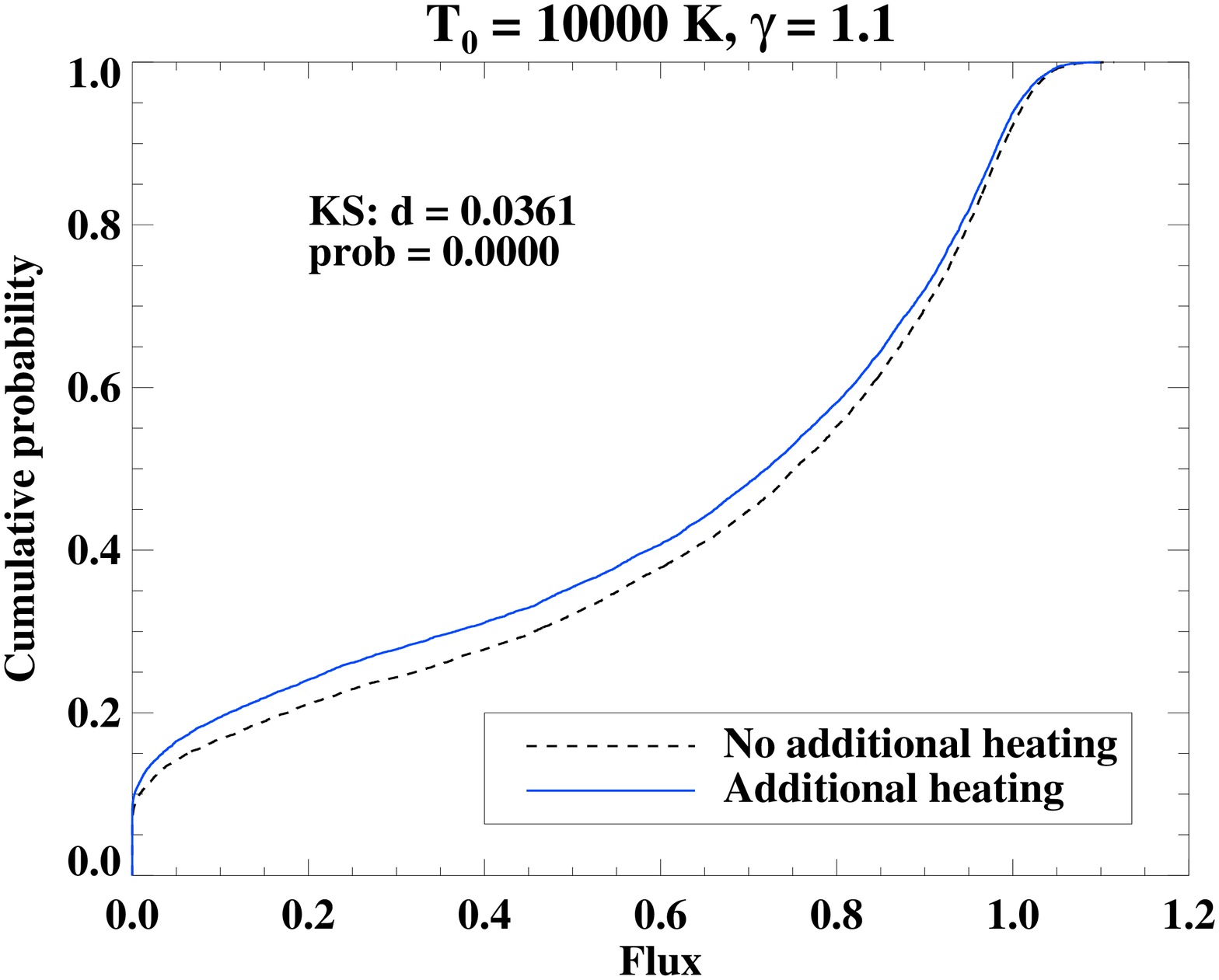} \includegraphics[scale=0.35, angle = 90]{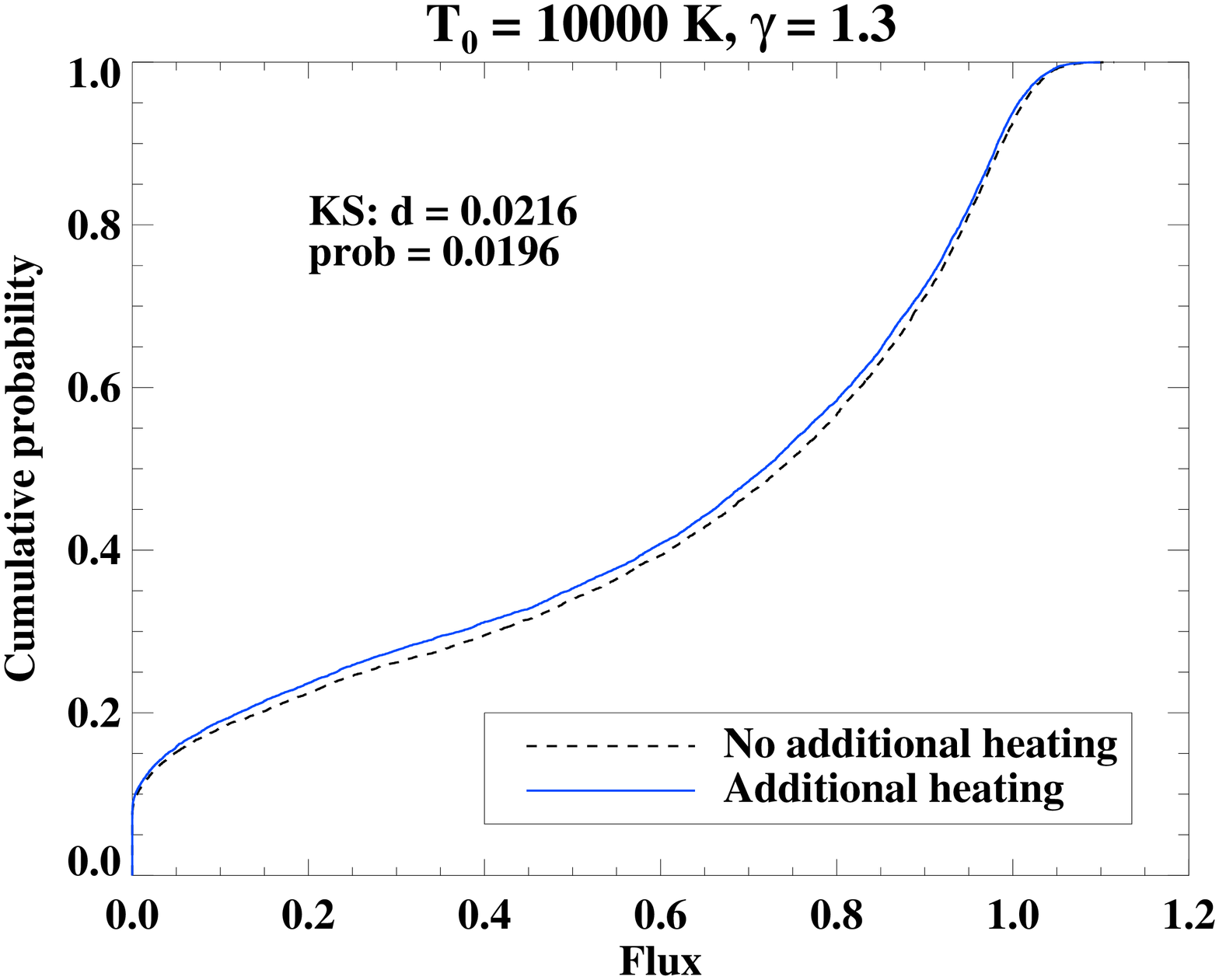} \includegraphics[scale=0.35, angle = 90]{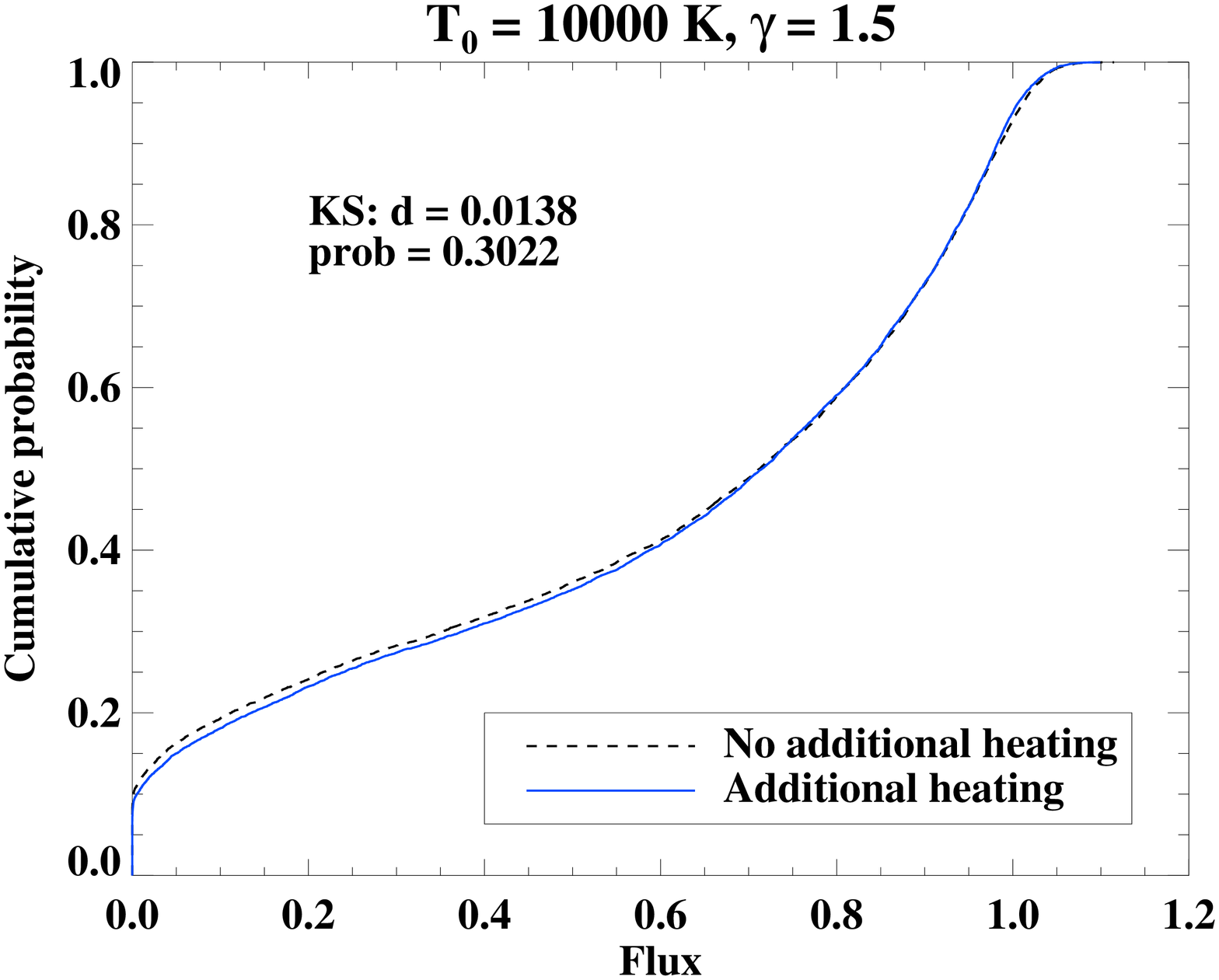}
  \caption{Comparison of the cumulative flux PDF of the spectrum with and without additional heating by the quasar for a sample of 20 lines-of-sight each having 512 pixels, drawn through the simulation box. The temperature at mean density is taken as 10$^4$ K with the slope being varied from 1.1 (top), 1.3 (middle) and 1.5 (bottom). The values of the KS statistics $d$ and prob are indicated on each panel. It can be seen that the distinguishability of the heated and non-heated spectra goes down as the slope of the equation of state is increased. With the smallest slope of 1.1, the spectra for the two cases are completely distinguishable even with 20 lines-of-sight.}
  \label{fig:eosfluxpdf1}
 \end{center}
\end{figure}

\begin{figure}
 \begin{center}
 \includegraphics[scale=0.35, angle = 90]{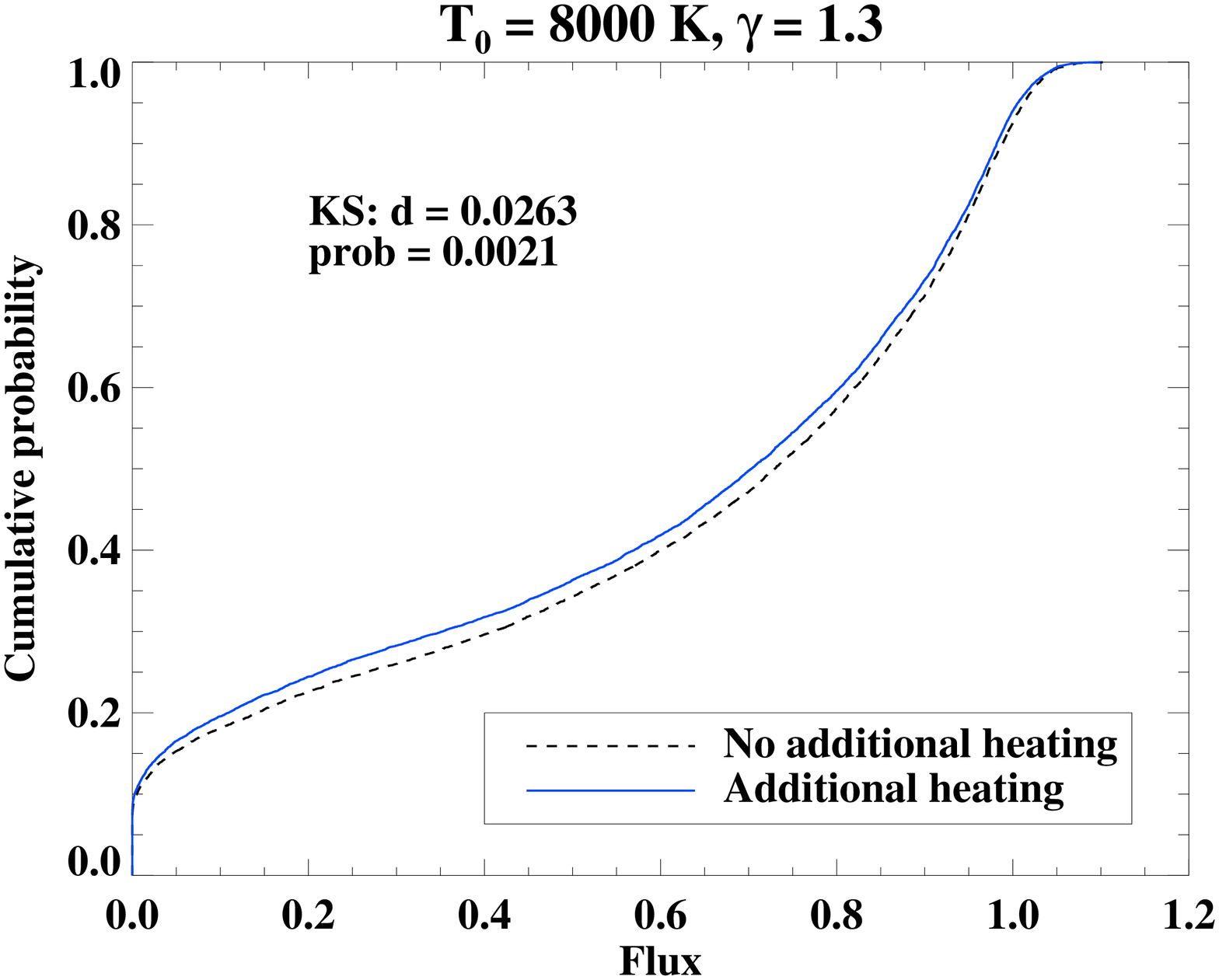} \includegraphics[scale=0.35, angle = 90]{fluxpdf_1_1.3_newseedlos.ps} \includegraphics[scale=0.35, angle = 90]{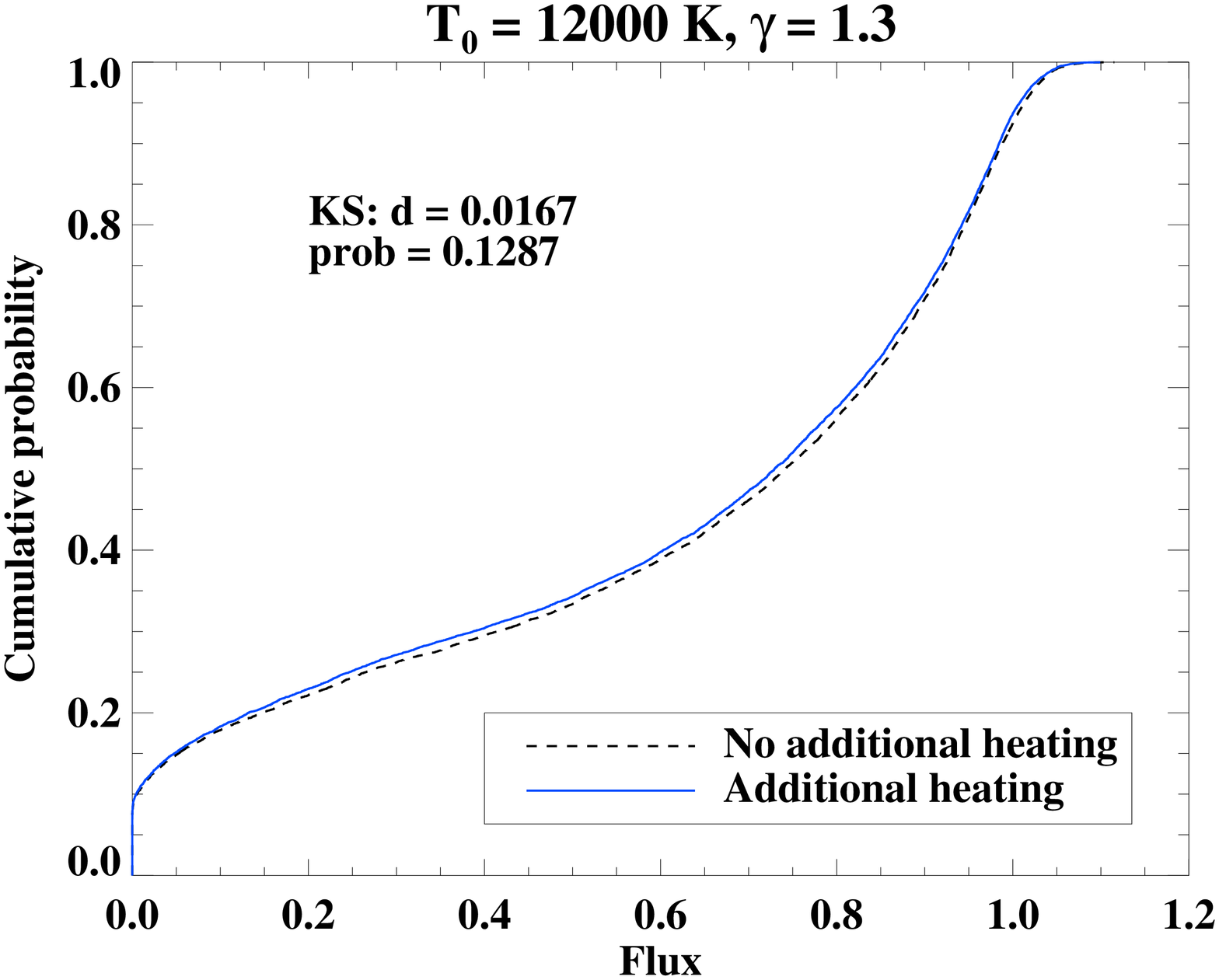}
  \caption{Same as Figure \ref{fig:eosfluxpdf1}, with the temperature at mean density being varied from 8000 K (top), 10000 K (middle) and 12000 K (bottom). The distinguishability of the heated and non-heated spectra goes down as the value of the temperature at mean density is increased. With the smallest temperature of 8000 K, the spectra for the two cases are distinguishable at the $99.79\%$ level even with 20 lines-of-sight.}
  \label{fig:eosfluxpdf2}
 \end{center}
\end{figure}

\begin{figure}
 \begin{center}
  \hskip-0.5in \includegraphics[scale = 0.36, angle = 90]{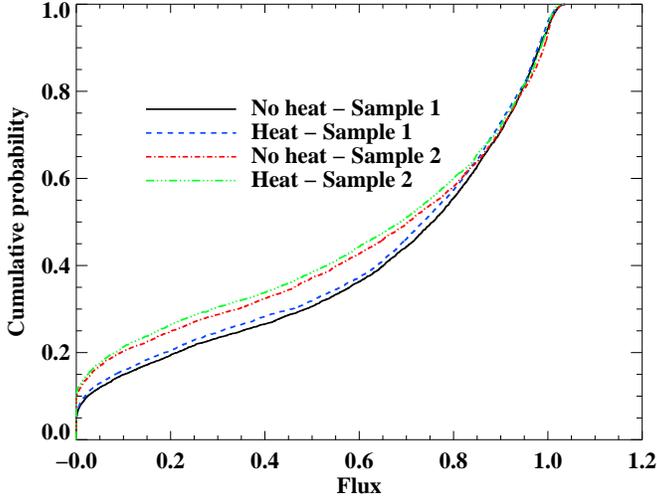}
  \caption{The cumulative probability distribution of the flux, for two samples each of control and heated spectra. Each sample comprises 10 lines-of-sight (5120 pixels). The blue dashed and green dot-dot-dot-dashed curves represent the heated samples and the black solid and red dot-dashed curves represent the control ones. It can be seen that the effect of the additional heating is within the cosmic variance of the individual samples.}
  \label{fig:cumulflux}
 \end{center}
\end{figure}

{We conclude that we are able to distinguish between the heated and control samples using 20 lines-of-sight and the flux PDF, and the extent of the distinguishability is sensitive to the initial parameters ($T_0$ and $\gamma$) of the equation of state. However, among these 20 lines-of-sight, we find that the statistical difference in the inferred flux PDF due to cosmic variance is greater than the difference introduced by additional heating from the quasar. This is summarized in Fig. \ref{fig:cumulflux} where we have plotted the cumulative probability distribution for two subsamples each from the control and the heated distributions. Each sample comprises 5120 pixels (10 lines-of-sight). It can be seen that the effect of additional heating on the flux PDF is within the cosmic variance of the individual samples. Hence, we infer that the flux PDF alone is not very sensitive to the additional heating effect,  but may be more sensitive to the \HI\ ionizing radiation, which is the purpose for which it is traditionally used.
}

\subsubsection{Curvature statistics}
\label{sec:curvature}

The flux PDF statistic points to a connection between the heating effect and the initial equation of state. However, the difference is within the individual cosmic variance of the samples, making it difficult for the technique to be used in practice to identify a given spectrum as being ``heated'' or not. In order to address this effect and also to isolate the effect of the additional heating from the ionization information (both of which are captured in the flux), we consider here a alternative statistic, to characterize the spectra. In the literature, this has been done in several ways:   (a) by using the $b$-distribution from Voigt profile fitting to the mock spectra \citep[e.g.][]{bolton12}, (b) by using wavelets \citep[e.g.][]{theuns2000} or (c) by using the curvature parameter \citep[e.g.][]{becker11}. Unlike in the case of low-redshift Lyman-$\alpha$ forest absorption, the $b$ parameter need not be well constrained as one will not be able to use higher Lyman series lines. In this section, we explore the usage of the curvature parameter, to analyse the heating effect statistically. Following \citet{becker11}, the curvature parameter can be defined as:

\begin{equation}
 \kappa = \left|\frac{F''}{(1 + (F')^2)^{3/2}}\right|
\end{equation} 
where $F$ is the normalized flux\footnote{Our $\kappa$ corresponds to $|\kappa|$ of \citet{becker11}.}.  The binned average of the curvature at a given flux, together with simulations, are used to measure the IGM temperature without resorting to Voigt profile fitting techniques by \citet{becker11}. As pointed out by these authors, the denominator of the above expression is essentially unity and hence only the double derivative of the flux contributes to the curvature. We follow \citet{becker11} where the flux (and all its derivatives) are measured with respect to the velocity grid in km/s. We evaluate the curvature parameter for both, the control and the heated spectra. In addition to the KS statistic for the flux, described in the previous subsections, we now also use the KS statistic for the $\kappa$ distribution and use the two dimensional KS statistic to compare the joint flux-$\kappa$ distributions. In this way, the effect of the additional heating may be quantified. 

We begin by calibrating the effect of the curvature statistic. To do this, we consider the fiducial equation of state, having parameters $T_0 = 10^4 \ \rm{K}, \gamma = 1.3$, and a single line-of-sight (512 pixels). We first generate noise-free spectra along the line-of-sight for both ``control'' and ``heated'' cases, and compute the curvature values for both of these. Noise is then added to both the control and heated samples, and the curvature values are again computed. Now, the control and the heated samples are statistically compared (using the KS test) with respect to the flux PDF, the curvature, and the joint flux-$\kappa$ distributions for both the cases, i.e. with and without noise added to the spectra.

 We find that when no noise is added to either the ``control'' or the ``heated'' spectra, then the three KS probabilities are 0.752 (for flux PDF alone), 0.002 (for $\kappa$ alone\footnote{Here, and in what follows, we disregard the pixels having flux values greater than 0.9 or less than 0.1, for all curvature statistics. This is done following \cite{becker11}, to avoid both, saturated pixels at low flux as well as uncertainties in the curvature values at high flux.}) and 0.021 (for the 2d KS test). This confirms that the curvature parameter is far better able to distinguish between the heated and the control samples than the flux PDF. This is to be expected since the curvature parameter directly captures the effect of thermal broadening.
 
On the other hand, when noise is added to both the ``control'' and ``heated'' spectra, then the above three probabilities become 0.316 (for flux PDF alone), 0.768 (for $\kappa$ alone) and 0.529 (for the 2d KS test). These values (also summarized in Table \ref{table:halos}) indicate that the curvature statistic is strongly influenced by the noise in the spectrum, which washes out the distinguishability of the control and the heated spectra. This has also been noted previously by \citet{becker11}.

\begin{table}
\begin{center}
    \begin{tabular}{ | c | c | c | p{5cm} |}
    \hline
     & No noise & With noise  \\ \hline
     Flux & 0.752 & 0.316 \\ 
     $\kappa$ & 0.002 & 0.768 \\
     2d KS test   & 0.021  & 0.529 \\ \hline
    \end{tabular}
\end{center}
\caption{This table indicates the KS test probabilities for the non-noise added and the noise added spectra. The KS test is performed between the control and the heated samples of 512 pixels each. The last row indicates the probability values for the two-dimensional KS test of the flux-$\kappa$ joint distribution. It can be seen that noise significantly affects the value of prob for the curvature statistic.}
 \label{table:halos}
\end{table}
 
\begin{figure}
 \begin{center}
  \hskip-0.4in \includegraphics[scale=0.35, angle = -90]{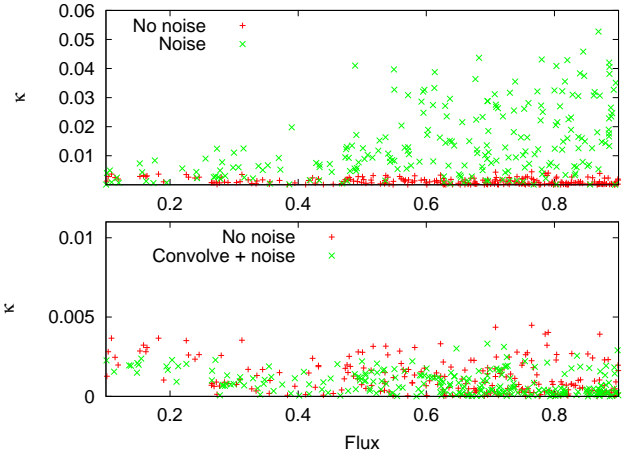}
  \caption{The top panel shows the 2D scatter plots of the flux-$\kappa$ distribution in the non-noise added (ideal) case (red plus signs), and the noise added case (green crosses). The distributions are significantly different. In the bottom panel, the non-noise added (ideal) distribution (red plus signs) is shown along with the  noisy spectrum convolved with a 10 km/s filter (green crosses). The figure shows that it is indeed possible to approach the ideal 2D distribution when the noise is convolved out with a smoothing velocity.}
  \label{fig:convnoise}
 \end{center}
\end{figure}

\begin{figure*}
 \begin{center}
   \includegraphics[scale=0.3, angle = 90]{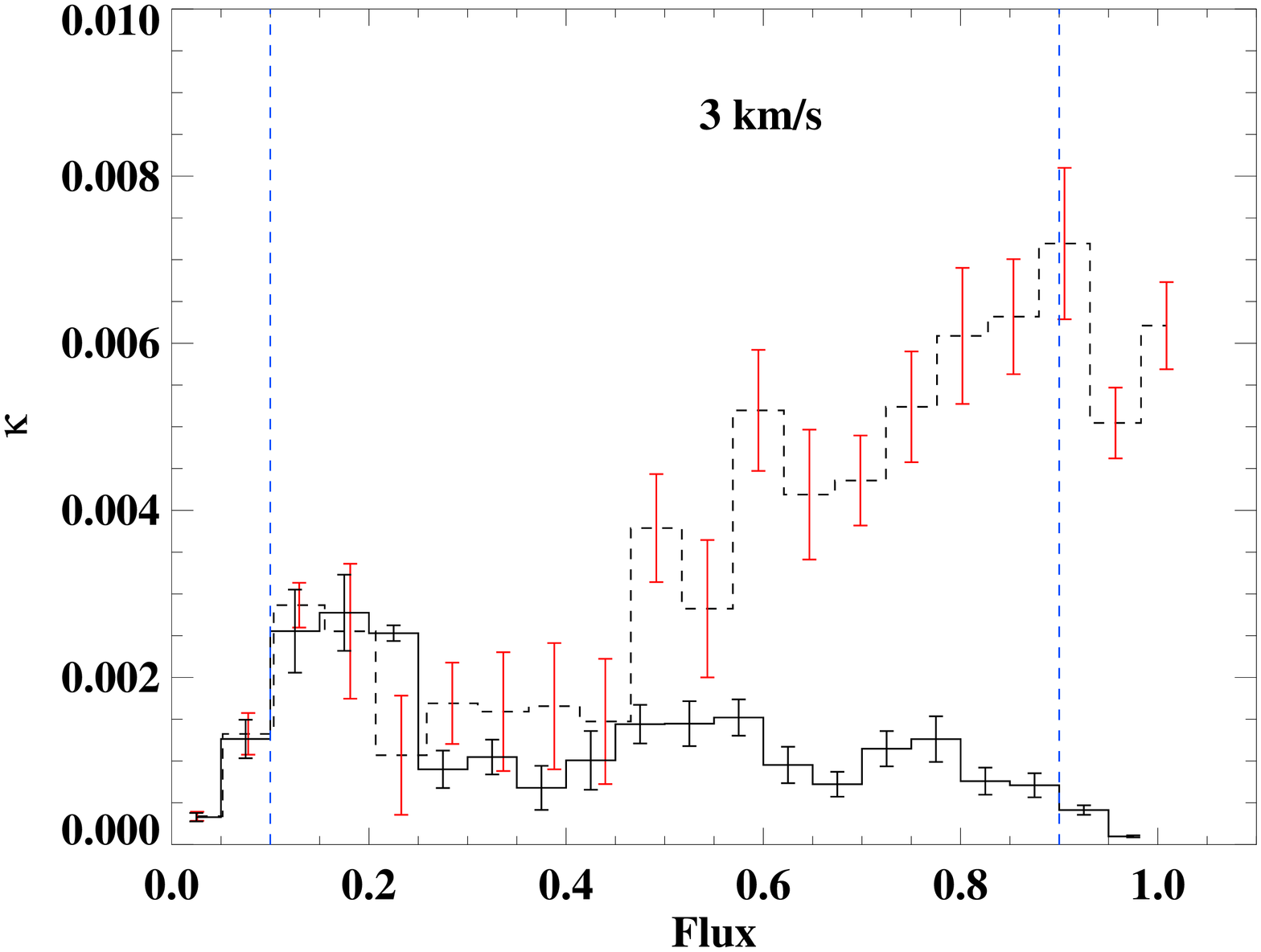} \includegraphics[scale=0.3, angle = 90]{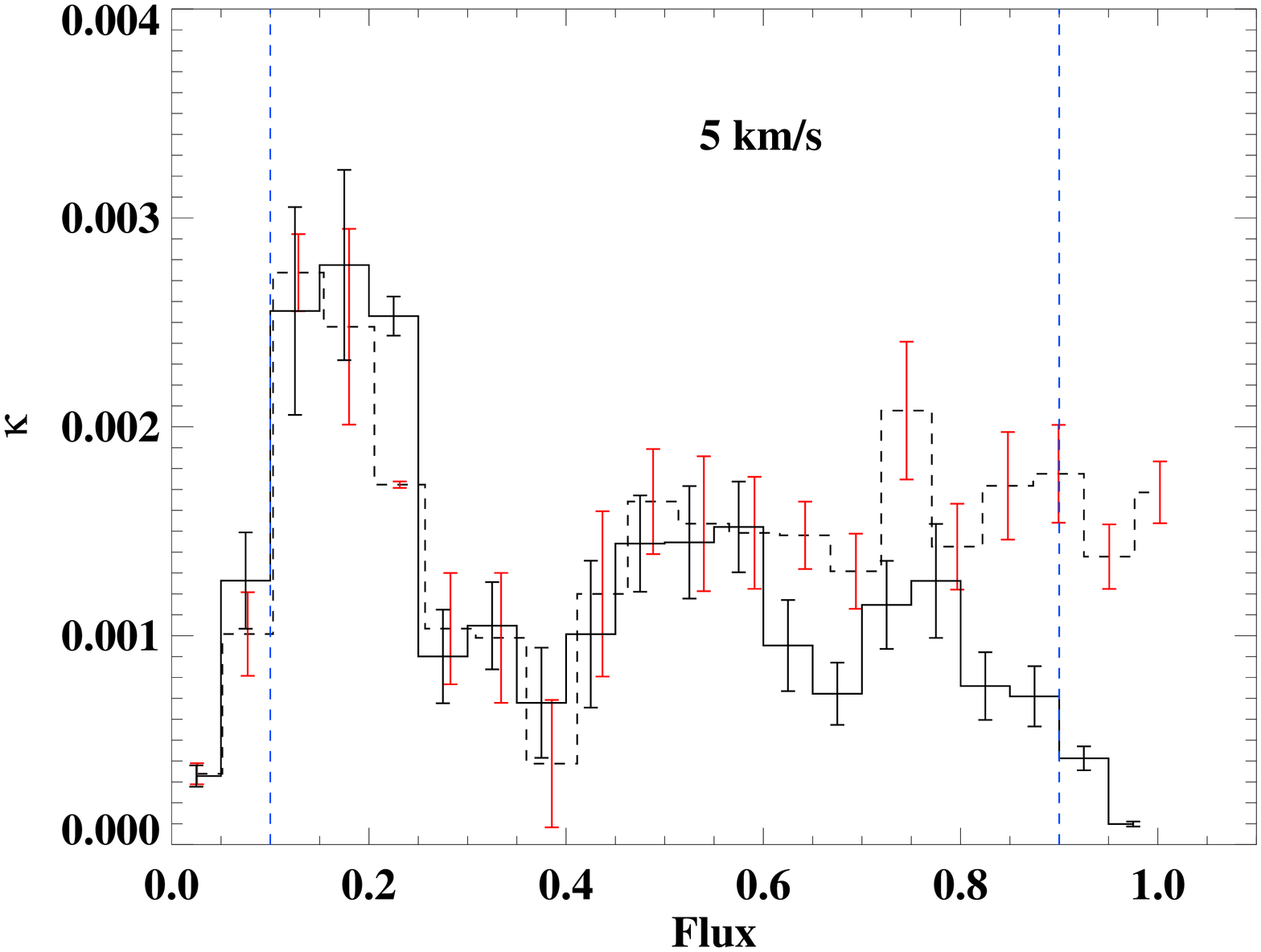} \includegraphics[scale=0.3, angle = 90]{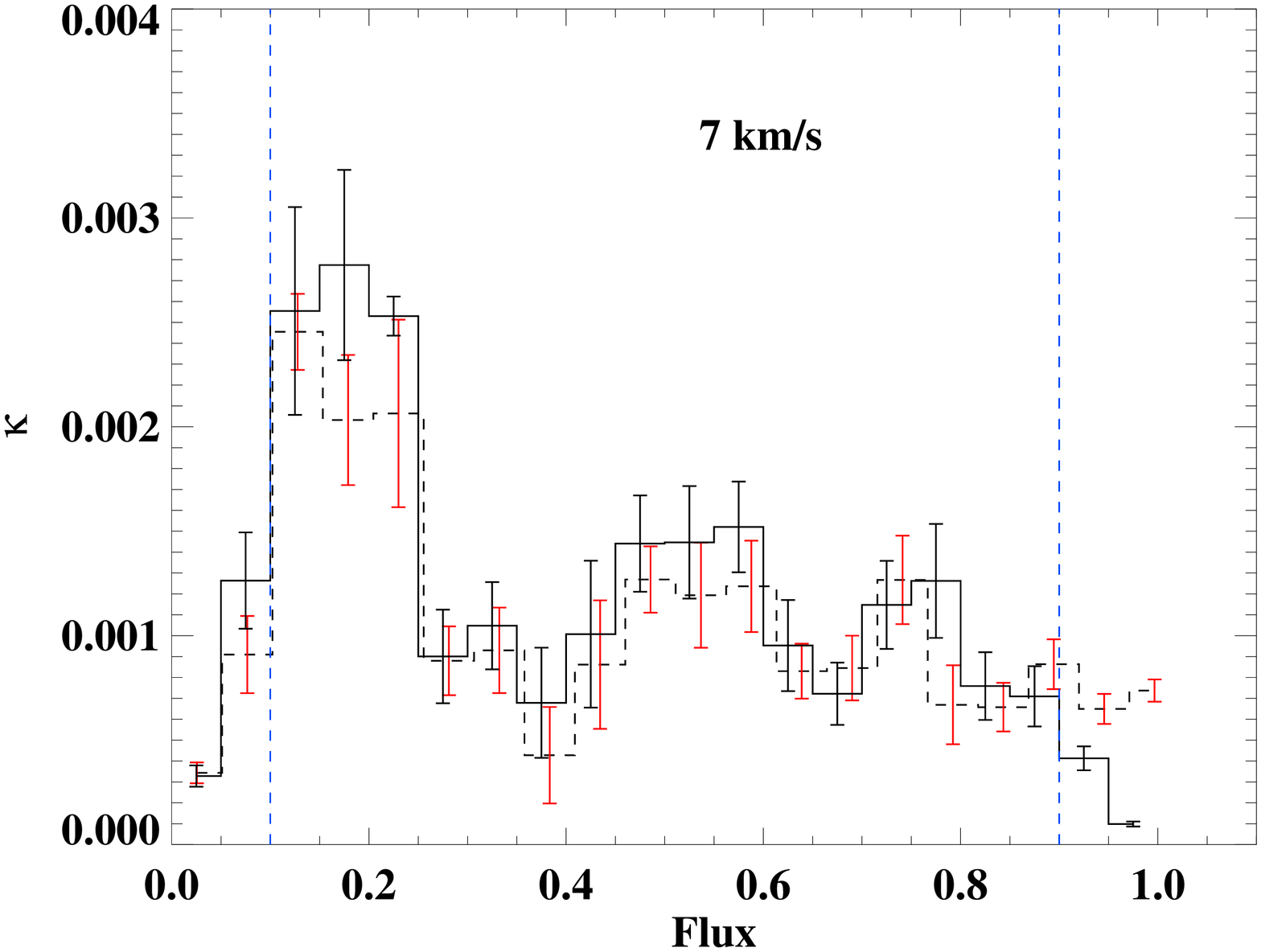}  \includegraphics[scale=0.3, angle = 90]{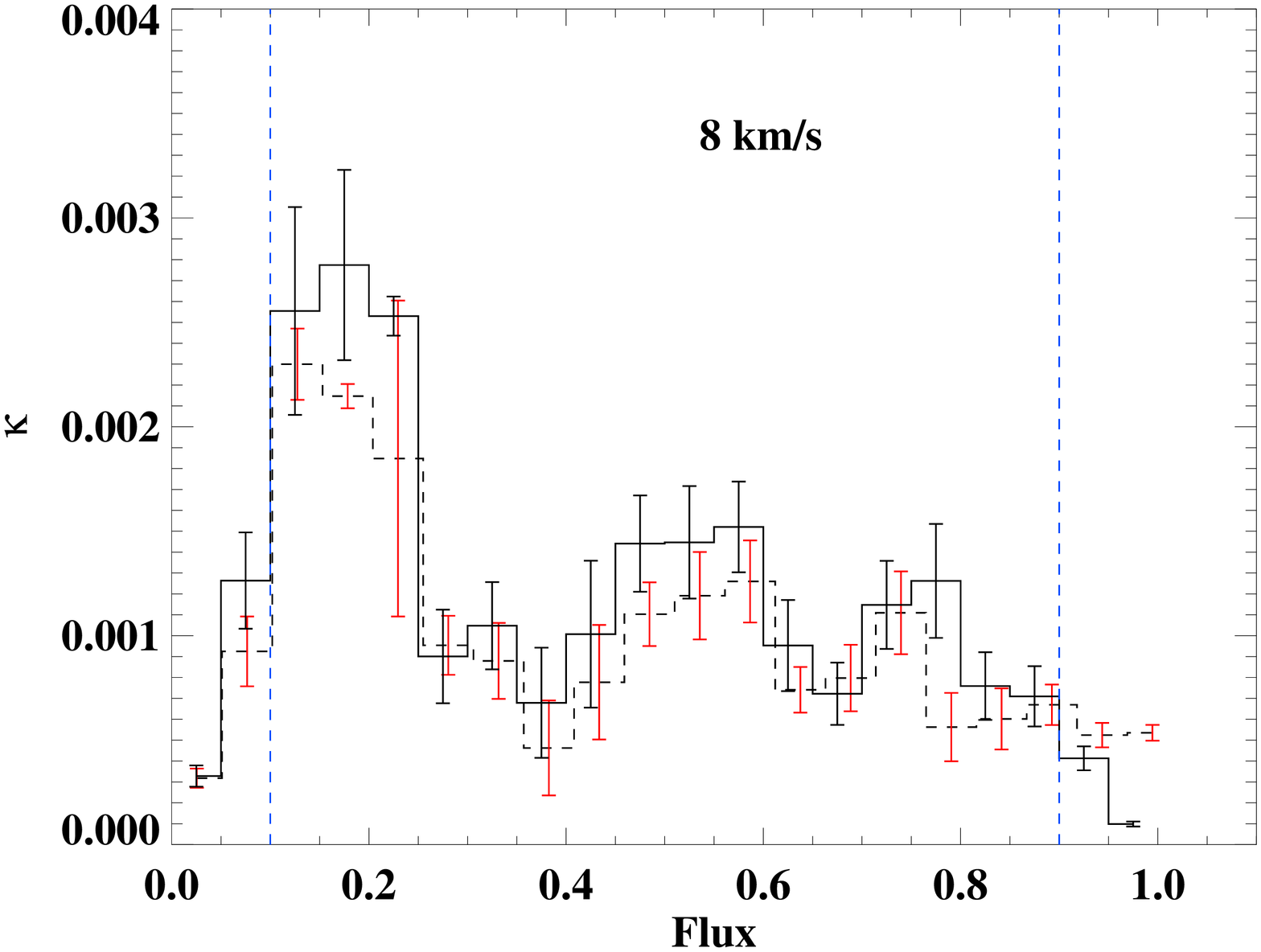}
  \caption{{The average $\kappa$ and the associated error in different flux bins are plotted versus flux (for the control spectrum). These plots show the approach of the convolved flux-$\kappa$ distribution (dashed lines) to the non-noise added (ideal) distribution (solid lines) using different smoothing velocities, 3 km/s, 5 km/s, 7 km/s and 8 km/s from top left to bottom right. The blue dotted lines indicate the limits of the range in flux used for all the curvature statistics ($0.1 \leq \rm Flux \leq 0.9$).  It can be seen that the smoothing velocity of 7 km/s (bottom left) most closely resembles the ideal distribution.}}
  \label{fig:diffvel}
 \end{center}
\end{figure*}

\begin{figure*}
 \begin{center}
  \includegraphics[scale=0.5, angle = -90]{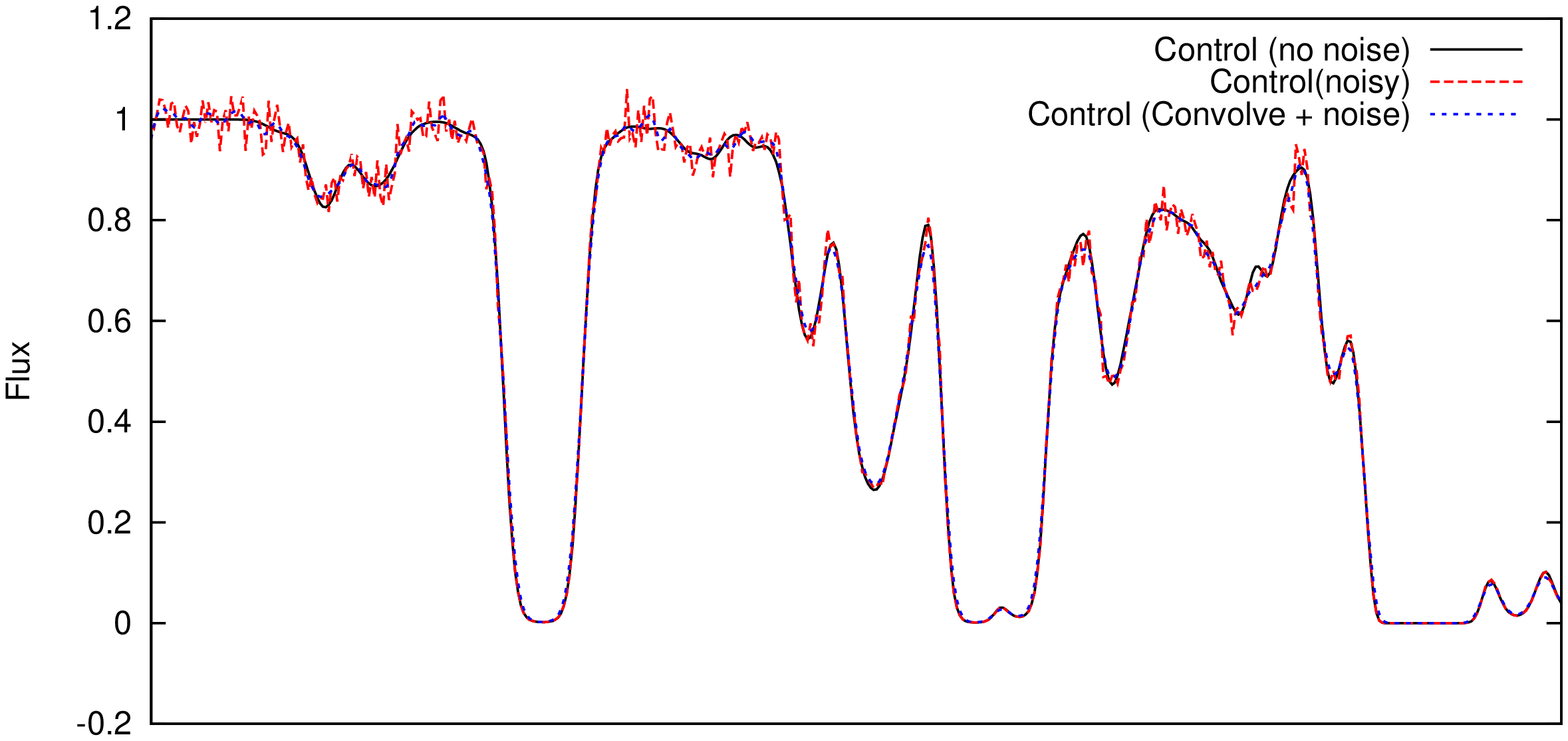} \vskip-0.6in
  \includegraphics[scale=0.5, angle = -90]{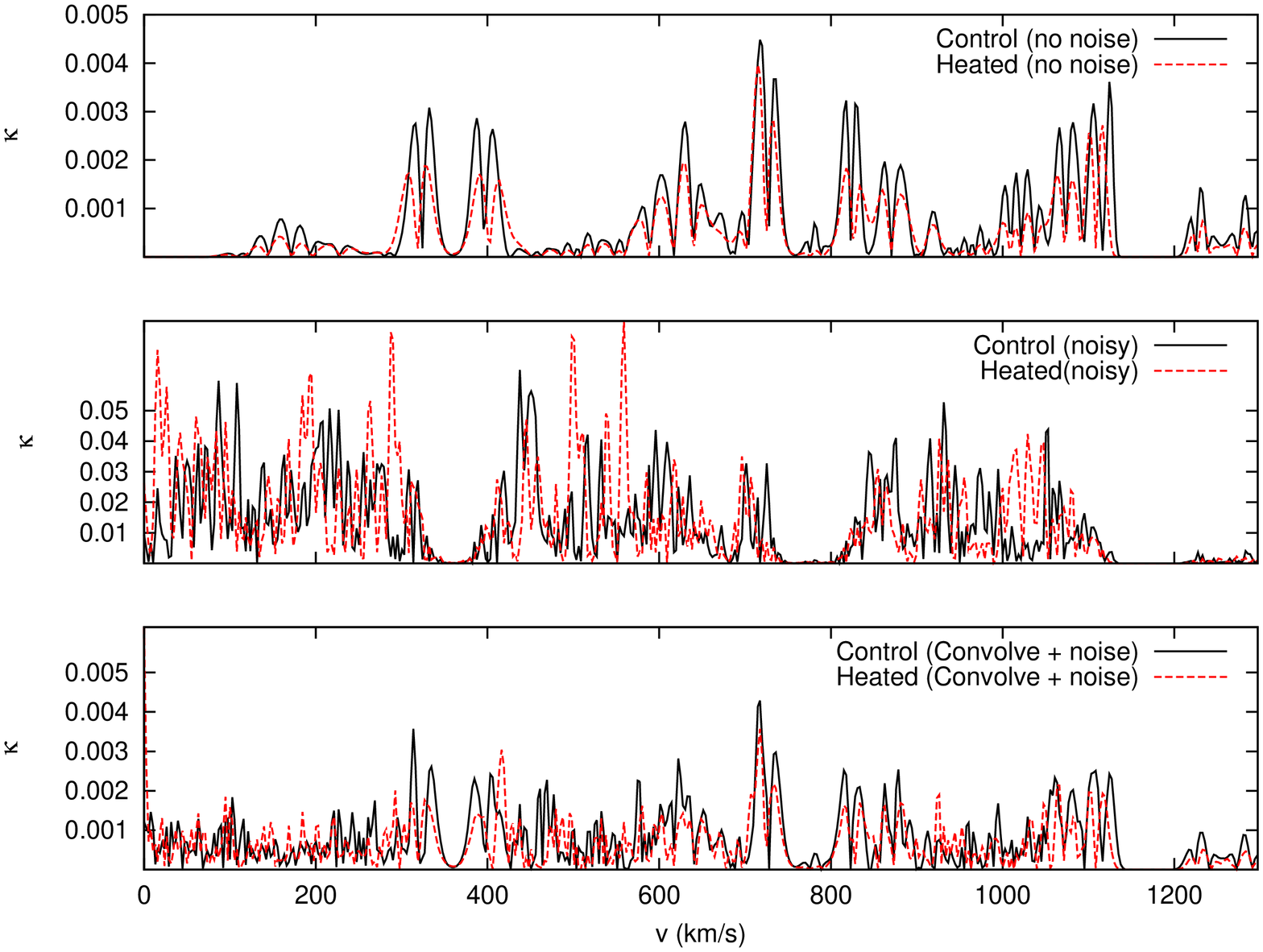}
  \caption{\textit{Top panel}: The (control) quasar spectrum for the three cases : no noise added, noise added and noise convolved with the Gaussian filter of 7 km/s. It may be seen that the convolution with the Gaussian filter closely approximates the ideal distribution. \textit{Lower three panels}: The curvature parameter as a function of pixel for three cases from top to bottom : no noise added, noise added, and noise convolved with the Gaussian filter of 7 km/s, for both the heated and the control samples along a line-of-sight.}
  \label{fig:curvmulti}
 \end{center}
\end{figure*}

Since the noise significantly dominates the curvature statistic, in order for the efficient usage of the curvature statistic, it is important to smooth the noisy spectrum before applying this statistic. In \citet{becker11}, this is achieved by fitting the raw spectra with a smoothly varying $b$-spline and the curvature is computed from the smoothed spectra. In this work, we convolve the noisy spectra with a Gaussian filter having a specific smoothing velocity width and vary the width until the convolved spectrum best matches the ideal, non-noise added spectrum. The results of this exercise are  illustrated in Figs. \ref{fig:convnoise} and \ref{fig:diffvel}. In Fig. \ref{fig:convnoise}, the top panel shows the 2D scatter plot of the $\kappa$-flux joint distribution for the control sample, with and without noise added to the spectrum. The bottom panel shows the noisy 2D distribution convolved with a Gaussian smoothing filter of 10 km/s, compared to the noise-free distribution. The figure shows that the noise is efficiently convolved out by smoothing with the Gaussian filter, since the convolved scatter plot closely resembles the original, non-noise added plot. We now fine-tune the value of the smoothing velocity until the  convolved distribution most closely matches the ideal non-noise added distribution, and plots for different smoothing velocities of 3, 5, 7 and 8 km/s are in Figure \ref{fig:diffvel}. It is seen that a smoothing velocity of 7 km/s most closely matches the non-noise added distribution and hence we adopt it for the subsequent analysis. This is also apparent from the plot in Fig. \ref{fig:curvmulti} which illustrates the pixel dependence of the flux and the curvature parameter for the three cases : no noise, noise added, and noise convolved with the 7 km/s Gaussian filter. We also note that the curvature parameter values we obtain are consistent (at the same order-of-magnitude) with those in \citet{becker11}\footnote{As an aside, we have found that smoothing with a moving boxcar distribution for different boxcar widths does not produce the systematic effects noted in Fig. \ref{fig:diffvel} and hence, the convolution with the Gaussian filter is preferred over the moving boxcar to smooth the distribution.}
.

We now vary the equation of state, and the 2d KS test between flux and $\kappa$ for 512 pixels (1 line-of-sight) yields the values in the second column of Table \ref{table:2dprobeos}. It can be seen that the trend of greater distinguishability with smaller $T_0$ and $\gamma$, which we found for the flux PDF case, is reproduced for the case of the curvature statistic as well. The curvature statistic can effectively distinguish between the control and heated spectra for different equations of state even with a sample of 512 pixels (a single line-of-sight). The prob values for a sample of five lines-of-sight are also provided in the last column of Table \ref{table:2dprobeos}. This shows that the distinguishability of the samples crosses the 90\% level with a sample of 5 sightlines (equivalent to using two quasar spectra) for all equations of state under consideration. If we use 20 lines-of-sight, the control and heated spectra are completely distinguishable (to less than about one part in $10^{8}$) for all equations of state under consideration.

\begin{table}
\begin{center}
    \begin{tabular}{ | c | c | c | p{5cm} |}
    \hline
     $T_0$, $\gamma$ & 2d KS prob   & 2d KS prob \\
                    & (1 line-of-sight) &  (5 lines-of-sight) \\ \hline
     10$^4$ K, 1.1 & 0.067 &  3.326 $\times 10^{-9}$ \\ 
     10$^4$ K, 1.3 & 0.146 &  8.955 $\times 10^{-5}$\\
     10$^4$ K, 1.5 & 0.801  &  0.093\\
     0.8 $\times$ 10$^4$ K, 1.3 & 0.071  &  1.936 $\times 10^{-9}$ \\
     1.2 $\times$ 10$^4$ K, 1.3  & 0.323  &   0.016\\ \hline
    \end{tabular}
\end{center}
\caption{This table indicates the two-dimensional KS test probabilities of the flux-$\kappa$ joint distribution for different equations of state with a sample of 512 pixels (1 line-of-sight) and 2560 pixels (5 lines-of-sight). The KS test is performed between the control and the heated samples. It can be seen that the distinguishability of the samples decreases as $T_0$ and/or $\gamma$ are increased, quantifying the dependence of the additional heating effect on the initial equation of state. Note that all the background photoionization rates are fixed at the HM12 values.}
 \label{table:2dprobeos}
\end{table}
 
\begin{figure}
 \begin{center}
  \hskip-0.4in \includegraphics[scale=0.35, angle = 90]{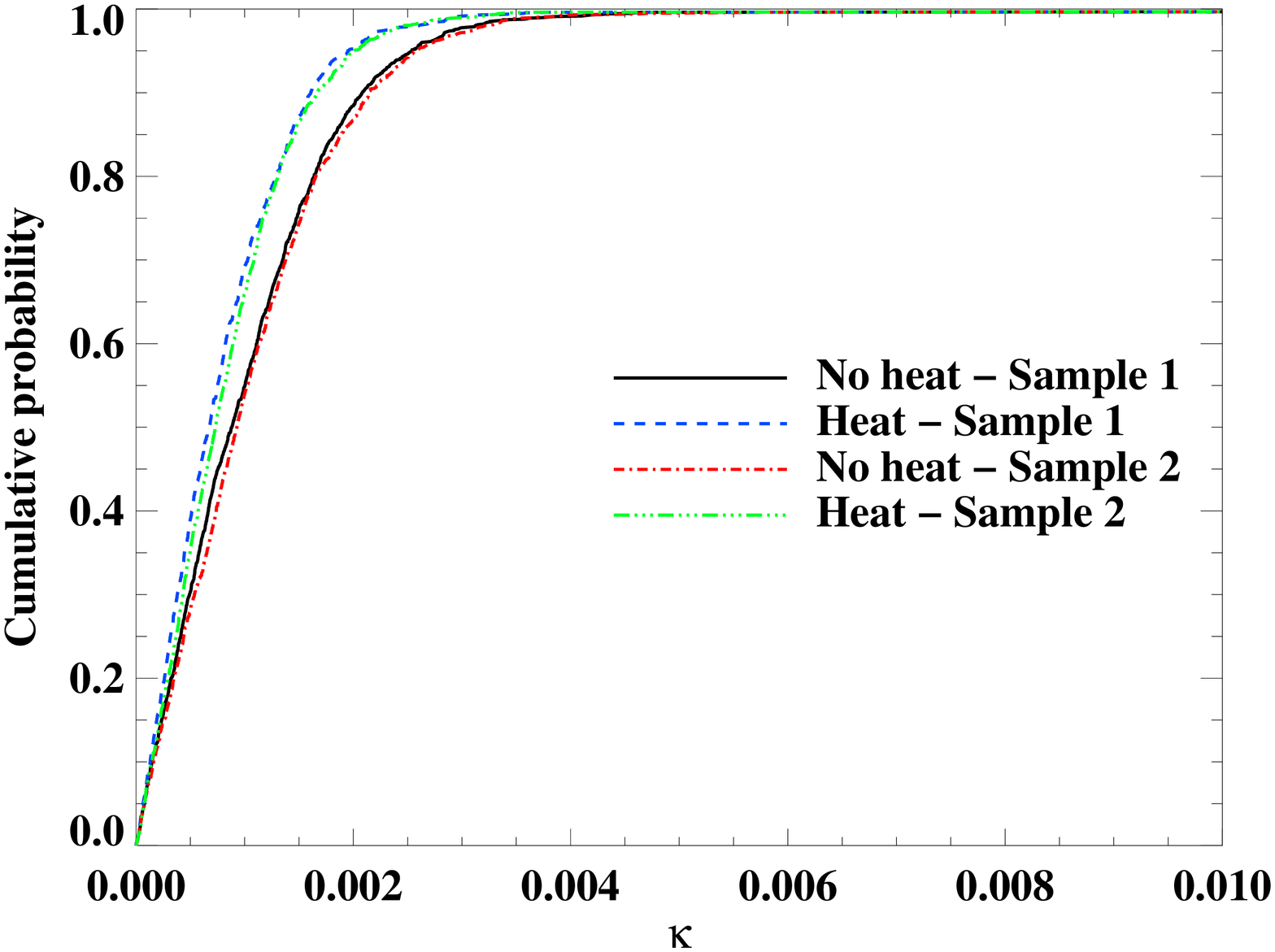}
  \caption{The cumulative probability distribution of the curvature statistic, $\kappa$, for two samples each of control and heated spectra. Each sample comprises 10 lines-of-sight (5120 pixels). The blue dashed and green dot-dot-dot-dashed curves represent the heated samples and the black solid and red dot-dashed curves represent the control ones. It can be seen that the effect of the additional heating is well above the cosmic variance of the individual samples. This figure may be compared with Fig. \ref{fig:cumulflux} where the opposite effect was noted.}
  \label{fig:cumulative}
 \end{center}
\end{figure}

In order to explore the extent of the effect of cosmic variance on our results, we consider now our fiducial equation of state and compare the cumulative probability distributions of the curvature statistic for two control subsamples, each of 10 sightlines, and two ``heated'' subsamples, each again of 10 sightlines. The resulting plot is shown in Fig. \ref{fig:cumulative}. The blue dashed and green dot-dot-dot-dashed curves represent the heated samples and the black solid and red dot-dashed curves represent the control samples. It may be clearly seen that the heating effect is well above the ``cosmic variances'' of the individual samples; this figure may be compared to the previous Fig. \ref{fig:cumulflux} where the opposite effect was noted. Hence, we conclude that the curvature statistic will be able to distinguish the ``non-heated'' and ``heated'' spectra over and above their internal cosmic variance even when we use a sample size as limited as what is available today. 

\subsection{Dependencies on single-step reionization by Population III stars}
\label{sec:popiii}

In the preceding sections, we have statistically quantified the dependence of the heating effect on the equation of state parameters ($T_0$ and $\gamma$). In standard two-step reionization scenarios, these two parameters may be mapped to the redshift of hydrogen reionization, and the associated IGM temperature at that redshift. In this section, we briefly consider the effects of our study on constraining single-step models of reionization.

In Sec. \ref{sec:validepen}, we illustrated the effects of changing the $\Gamma^{\rm bg}_{\rm HeII}$ photoionization rate on the temperature-density distribution, for different initial values of the normalization of the equation of state, $T_0$. We also indicated which combinations of these two parameters produced results which were consistent with those measured in the near-zones of the $z \sim 6$ quasars \citep{bolton12}. It was found that when the $\Gamma^{\rm bg}_{\rm HeII}$ was small (or when the initial $x_{\rm HeII}$ was high), $T_0$ showed the maximum increase with no apparent change in $\gamma$. However, as the $\Gamma^{\rm bg}_{\rm HeII}$ became higher, while the increase in temperature was moderate, we found that the equation of state became steeper (i.e. $\gamma$ became higher). As the curvature statistics uses the whole spectra, it should be sensitive to changes in both $T_0$ and $\gamma$. Therefore, we now discuss how the detectability of the heating effect depends on the initial value of $x_{\rm HeII}$.
This, in turn, can be connected to early reionization of both \HI\ and \HeII\ by massive stars in single-step models \citep{venkatesan2003, wyithe2003, trc2005, trc2006}. In the single-step model of reionization, Population III stars reionize both \HI\ and \HeII\ at redshifts $z > 6$. In some single-step models \citep{venkatesan2003}, the fraction of helium in \HeIII\ may hence reach about $60 \%$ by $z \sim 5.6$, which translates into $x_{\rm HeII}$ being only of the order of $\sim 0.4$. 

In order to investigate the effect of a lower initial $x_{\rm HeII}$ in the quasar near-zone, we consider different values of the metagalactic background $\Gamma^{\rm bg}_{\rm HeII}$ which translates into varying the initial \HeII\ fraction, $x_{\rm HeII}$, and investigate the detectability of the additional heating to the variation of $x_{\rm HeII}$. The fiducial equation of state parameters, $T_0 = 10^4$ K, and $\gamma = 1.3$ are used in this study. For each value of $\Gamma^{\rm bg}_{\rm HeII}$ which we consider, we generate ``control'' and ``heated'' spectra, then these two samples are compared using the 2d Kolmogorov-Smirnov statistic. The results are indicated in Table \ref{table:2dprobxheii}. 

The table shows that the effect of the additional heating is more apparent if the initial fraction of $x_{\rm HeII}$ is greater. This is to be expected from the qualitative indications in Fig. \ref{fig:constraints1}, since a greater $x_{\rm HeII}$ fraction leads to a higher final (heated) temperature, and hence a greater difference between the control and the heated samples. The argument may be reversed to provide constraints on the metagalactic \HeII\ background required before the quasar is turned on, in order for the the additional heating effect to be detected at a particular level. For example, with all other parameters being equivalent, if the additional heating effect is to be detected with greater than 75 \% confidence, then the initial \HeII\ fraction in the vicinity of the quasar is constrained to $\gtrsim 0.74$, which, in turn, constrains the $\Gamma^{\rm bg}_{\rm HeII}$ to $\lesssim 10^{-17}$. Consequently, we infer that in single-step models of reionization where the $x_{\rm HeII}$ in the quasar vicinity takes very small values, 
the additional heating effect may be considerably less detectable than in two-step models, which allow for a greater \HeII\ fraction in the quasar near-zone.

\begin{table}
\begin{center}
    \begin{tabular}{ | c | c | c | p{5cm} |}
    \hline
     $\Gamma^{\rm bg}_{\rm HeII} (\rm{in \ units \ of} \ HM12)$ & $x_{\rm HeII} (\rm initial)$ & 2d KS prob  \\ \hline
     $10^{4}$ & 0.040 & 0.501  \\ 
     $10^{3}$ & 0.260 & 0.291  \\
     $10^{2}$ & 0.741 & 0.210   \\
     $10$ & 0.963 & 0.148    \\
     $1$ & 0.996 & 0.146    \\ \hline
    \end{tabular}
\end{center}
\caption{This table indicates the two-dimensional KS test probabilities of the flux-$\kappa$ joint distribution for different initial \HeII\ fractions with a sample of 512 pixels (1 line-of-sight). The KS test is performed between the control and the heated samples. It can be seen that the distinguishability of the samples decreases if the initial \HeII\ fraction is lower (or equivalently, if the \HeII\ metagalactic background is higher), thus quantifying the dependence of the additional heating effect on the initial \HeII\ fraction. In the above table, the initial equation of state is fixed at the fiducial value ($T_0 = 10000$ K, $\gamma = 1.3$.)}
 \label{table:2dprobxheii}
\end{table}

Hence, we have effectively probed the sensitivity of the curvature statistic to the initial \HeII\ fraction in the vicinity of the quasar. However, as we saw in Sec. \ref{sec:validepen}, the change in the \HeII\ fraction leads to both, a moderate increase in temperature as well as a steepening of the slope. In the preceding subsections while discussing the curvature statistics, we have kept the value of the $\Gamma^{\rm bg}_{\rm HeII}$ fixed at the HM12 value. This, as we have seen in Sec. \ref{sec:validepen}, leads to a shift in the overall equation of state with no apparent change in slope. Hence, by performing the KS test in the previous subsections, we have equivalently captured the sensitivity of the curvature statistics to a change in the overall normalization, and seen that an overall normalization shift may be readily distinguished even with a sample of 512 pixels. Here, we briefly indicate the complementary effect, i.e. the sensitivity of the curvature statistic to a change in the slope alone. Note that this effect would not be captured in the previous tests with the curvature statistics, since the small (HM12) value of $\Gamma^{\rm bg}_{\rm HeII}$ considered therein, ensured that the slope change between the control and heated samples was negligible. We have seen in Fig. \ref{fig:changeingamma} in Sec. \ref{sec:validepen} that the maximum expected change in the initial $\gamma = 1.3$, for the lowest initial temperatures and \HeII\ fractions under consideration, is of the order of $\Delta \gamma = 0.1.$ We find that a sample of 512 pixels can distinguish $\Delta \gamma = \pm 0.1$ with only about 7\% - 39 \% confidence.
 If one also takes into account the observational and other sources of errors, we speculate that the allowed change ($\sim 0.1$) in $\gamma$ may be more difficult to detect statistically than the allowed change ($\sim 1000 - 5000$ K) in $T_0$. This also shows us that the curvature statistic is more sensitive to the detection of the change in the normalization than to the change in the slope of the equation of state.

\subsection{Influence of other effects}
\label{sec:othercont}
{

Here, we provide discussions of the other factors that may also influence the observed spectra in the quasar near-zones, and an analysis on their significance for this study.

\begin{figure}
 \begin{center}
 \hskip-0.5in \includegraphics[scale=0.35, angle = 90]{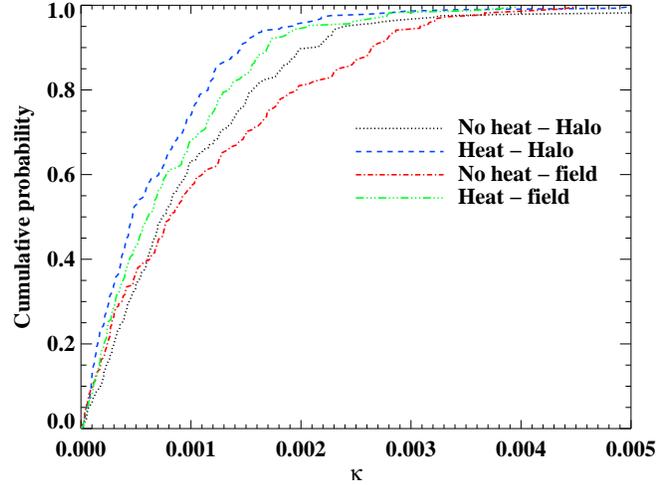}
  \caption{{Cumulative probability distribution along a sightline, for the curvature for the halo and field locations of the quasars. The median values of curvature are lower in the case of quasars residing in biased locations, corresponding to higher temperatures both for the initial and heated spectra.}}
  \label{fig:curvcomp}
 \end{center}
\end{figure}

\begin{enumerate}
 \item {Quasars in biased regions: } In order to explore the effects of locating quasars in biased regions, we extracted spectra by placing the quasar at the most massive halo (mass $\sim 1.49 \times 10^{11} h^{-1} M_{\odot}$) in the simulation box. The baryonic overdensities around the quasar are found to be $\sim 10-50$, and hence, on applying the equation of state, the initial temperatures are $\sim 20000 - 35000$ K. The curvature median values for biased locations of the quasar are slightly lower than those for unbiased locations, as plotted in Fig. \ref{fig:curvcomp}. 
 
 From the plots in Fig. \ref{fig:constraints1}, we speculate that there arises a degeneracy between the heating effects and the overdensities in the vicinity of the quasars when the quasars are modelled in biased locations. Temperatures of $T_0 \sim 20000$ K at redshift 6 arise in models where reionization occurs relatively late, $z_{\rm re} \lesssim 8$. Therefore, the density enhancement and consequent temperature enhancement in biased locations may lead to estimates of a later epoch of reionization, than if the quasars are modelled in unbiased regions. A caveat to this discussion is the assumption of the initial equation of state being valid even at large overdensities of 10-50 that arise in biased regions. A detailed treatment including the effects of shocks, etc. may be required to estimate the initial temperatures in these cases. However, this effect is expected to be minor, as indicated by the findings of \citet{raskutti2012}.

 \item {Three-dimensional effects and the environment: }
 Our simulations use a 1D treatment of radiative transfer. In reality, the quasar radiates in 3D with some finite opening angle. However, since the observations of quasar spectra and the Lyman-$\alpha$ forest are always along a line-of-sight or a set of several sightlines, the line-of-sight treatment of radiative transfer is adequate for producing simulated spectra and for the further statistical analyses. 
 
Recently, \citet{morselli2014} have detected the presence of galaxy overdensities in the environment of four $z \sim 6$ quasars. As stellar spectra are generally soft, there is negligible emission above 4 Ryd from galaxies, which is required for the ionization of \HeII. In other words, while the value of $\Gamma_{\rm HI}$ may be changed slightly, there is negligible contribution to $\Gamma_{\rm He II}$ from the galaxies. Since the dominant contribution to the heating effect comes from the ionization of \HeII, the heating effects and their detectability are influenced very little by the galaxies in the quasar environment. 
 \item {Variations in background HI ionizing flux:}
 We assume a uniform value of $\Gamma^{\rm bg}_{\rm HI}$ in the general IGM prior to the ``switching on'' of the quasar. Close to reionization, the value of $\Gamma^{\rm bg}_{\rm HI}$ may show spatial variations. However, our results do not change significantly since they are not sensitive to the actual value of $\Gamma^{\rm bg}_{\rm HI}$. Essentially, in the near-zone, the quasar ionizing flux dominates the background (by about a factor of 100 or more). Hence, small fluctuations in the initial $\Gamma^{\rm bg}_{\rm HI}$ are not expected to have a significant impact on the results and statistical analyses.

 \end{enumerate}

\section{Summary of main results}
 
In this paper, we used detailed hydrodynamical simulations to provide an analysis of the features associated with the heating due to the ionization of \HeII\ in the near-zones of high-redshift quasars, and their implications for constraining the epochs of \HI\ and \HeII\ reionization. 
Our main findings may be summarized as follows:

\begin{enumerate}
\item
We have seen that the measured temperature \citep{bolton12} in the quasar near-zones arises from a combination of two effects : the initial \HeII\ fraction in the quasar vicinity, and the normalization of the initial equation of state of the IGM. If the initial temperature at mean density is $\lesssim 8000$ K, the measured temperature in the quasar near-zones is higher than that expected for the allowed range of initial \HeII\ fractions ($x_{\rm He II} = 0.04 - 1$) in the quasar vicinity. This shows that the temperature measurement can be used to place constraints on (a) the epoch and temperature of hydrogen reionization, and (b) single-step models of reionization that predict the initial \HeII\ fraction.

\item We recover the expected linear relationship of $\Delta T_0$ increasing with the initial helium fraction $x_{\rm HeII}$. Akin to the $\Delta T- x_{\rm HeII}$ relation discussed in the literature \citep{furlanetto2008}, we also demonstrate a $\Delta \gamma - x_{\rm He II}$ relation, which shows a decrease in $\Delta \gamma$ with increasing $x_{\rm HeII}$ and a flattening out at the lowest $x_{\rm HeII}$ values, thus illustrating the steepening of the equation of state with decrease in the \HeII\ fraction in the quasar vicinity. Observationally, this steepening effect, which persists even for high initial temperatures where $\Delta T_0$ is low, may also be used to constrain the near-zone \HeII\ fraction.  However, the maximum expected increase in the slope may be more difficult to detect observationally than the expected shift in the overall normalization.  

\item Optical depth effects are coupled to the propagation of the ionization front in the radiative transfer, so that we obtain a handle on the extent of the near-zone of \HeIII, where the additional heating is expected to contribute significantly. If the quasar age is $\sim 100$ Myr, more than 80\% of the \HI\ proximity zone is heated in 78\% of the sightlines. The heated fraction of the \HI\ proximity zone is only about 30\% - 35\% for quasar lifetimes of $\sim 10$ Myr. This indicates that including the entire extent of the \HI\  proximity zone for the temperature enhancement may result in some dilution of the statistics when the quasar lifetimes are short. However, considering the entire  proximity zone of \HI\ is a valid approximation if the quasar lifetimes are longer, $\gtrsim 100$ Myr. This is also the timescale for the saturation of the heating effect, making it fairly independent of distance.

\item We have quantified the effect of additional heating by using the flux PDF and curvature statistics to compare the real spectra to the simulated spectra without heating. We have noted that the sensitivity of the curvature statistic to the noise in the spectra may be effectively removed by smoothing with a Gaussian filter with a velocity width of 7 km/s. Both these statistics indicate that a higher value of $T_0$ and/or $\gamma$ leads to less detectability of the effect of additional heating. This connects the additional heating due to \HeII\ reionization, to the epoch of hydrogen reionization.

\item We find that the curvature statistic provides far more effective distinguishability of the heating effect, which is over and above the cosmic variance of individual samples of 10 lines-of-sight each having 512 pixels (chosen to match the typical sample sizes available in observations of seven quasars at redshift $\sim 6$). We also find that the detectability of the heating effect is dependent on the initial \HeII\ fraction in the quasar vicinity, with a greater \HeII\ fraction leading to greater detectability.

\end{enumerate}
}

\section{Acknowledgements}
The research of HP is supported by the Shyama Prasad Mukherjee research grant of the Council of Scientific and Industrial Research (CSIR), India. The hydrodynamical simulations were performed using the Cetus and Perseus clusters of the IUCAA High Performance Computing Centre. HP thanks Jayanti Prasad and Vikram Khaire for helpful discussions. {We thank George Becker, James Bolton, Martin Haehnelt, T. Padmanabhan, Patrick Petitjean and David Syphers for useful comments on the manuscript. We thank the anonymous referee for helpful suggestions that improved the quality of the presentation.}

\bibliographystyle{mn2e} 
\def\aj{AJ}                   
\def\araa{ARA\&A}             
\def\apj{ApJ}                 
\def\apjl{ApJ}                
\def\apjs{ApJS}               
\def\ao{Appl.Optics}          
\def\apss{Ap\&SS}             
\def\aap{A\&A}                
\def\aapr{A\&A~Rev.}          
\def\aaps{A\&AS}              
\def\azh{AZh}                 
\def\baas{BAAS}
\def\jcap{JCAP}
\def\jrasc{JRASC}             
\def\memras{MmRAS}
\def\na{New Astronomy}
\def\nat{Nature}
\def\mnras{MNRAS}             
\def\pra{Phys.Rev.A}          
\def\prb{Phys.Rev.B}          
\def\prc{Phys.Rev.C}          
\def\prd{Phys.Rev.D}          
\def\prl{Phys.Rev.Lett}       
\def\pasp{PASP}               
\def\pasj{PASJ}
\def\physrep{Phys. Repts.}
\def\qjras{QJRAS}             
\def\skytel{S\&T}             
\def\solphys{Solar~Phys.}     
\def\sovast{Soviet~Ast.}      
\def\ssr{Space~Sci.Rev.}      
\def\zap{ZAp}                 
\let\astap=\aap
\let\apjlett=\apjl
\let\apjsupp=\apjs

\bibliography{mybib}

\begin{thebibliography}{79}
\expandafter\ifx\csname natexlab\endcsname\relax\def\natexlab#1{#1}\fi

\bibitem[{{Anninos} {et~al}\mbox{.}(1997){Anninos}, {Zhang}, {Abel}, \&
  {Norman}}]{abel}
{Anninos} P., {Zhang} Y., {Abel} T., {Norman} M.~L., 1997, \na, 2, 209

\bibitem[{{Assef} {et~al}\mbox{.}(2011){Assef}, {Kochanek}, {Ashby}, {Brodwin},
  {Brown}, {Cool}, {Forman}, {Gonzalez}, {Hickox}, {Jannuzi}, {Jones}, {Le
  Floc'h}, {Moustakas}, {Murray}, \& {Stern}}]{assef2011}
{Assef} R.~J. {et~al.}, 2011, \apj, 728, 56

\bibitem[{{Ba{\~n}ados} {et~al}\mbox{.}(2013){Ba{\~n}ados}, {Venemans},
  {Walter}, {Kurk}, {Overzier}, \& {Ouchi}}]{banados2013}
{Ba{\~n}ados} E., {Venemans} B., {Walter} F., {Kurk} J., {Overzier} R., {Ouchi}
  M., 2013, \apj, 773, 178

\bibitem[{{Becker} {et~al}\mbox{.}(2011){Becker}, {Bolton}, {Haehnelt}, \&
  {Sargent}}]{becker11}
{Becker} G.~D., {Bolton} J.~S., {Haehnelt} M.~G., {Sargent} W.~L.~W., 2011,
  \mnras, 410, 1096

\bibitem[{{Bolton} {et~al}\mbox{.}(2012){Bolton}, {Becker}, {Raskutti},
  {Wyithe}, {Haehnelt}, \& {Sargent}}]{bolton12}
{Bolton} J.~S., {Becker} G.~D., {Raskutti} S., {Wyithe} J.~S.~B., {Haehnelt}
  M.~G., {Sargent} W.~L.~W., 2012, \mnras, 419, 2880

\bibitem[{{Bolton} {et~al}\mbox{.}(2010){Bolton}, {Becker}, {Wyithe},
  {Haehnelt}, \& {Sargent}}]{bolton10}
{Bolton} J.~S., {Becker} G.~D., {Wyithe} J.~S.~B., {Haehnelt} M.~G., {Sargent}
  W.~L.~W., 2010, \mnras, 406, 612

\bibitem[{{Bolton} \& {Haehnelt}(2007{\natexlab{a}})}]{bolton07}
{Bolton} J.~S., {Haehnelt} M.~G., 2007{\natexlab{a}}, \mnras, 374, 493

\bibitem[{{Bolton} \& {Haehnelt}(2007{\natexlab{b}})}]{boltonhaehnelt07}
{Bolton} J.~S., {Haehnelt} M.~G., 2007{\natexlab{b}}, \mnras, 382, 325

\bibitem[{{Calverley} {et~al}\mbox{.}(2011){Calverley}, {Becker}, {Haehnelt},
  \& {Bolton}}]{calverley2011}
{Calverley} A.~P., {Becker} G.~D., {Haehnelt} M.~G., {Bolton} J.~S., 2011,
  \mnras, 412, 2543

\bibitem[{{Cassata} {et~al}\mbox{.}(2013){Cassata}, {Le F{\`e}vre}, {Charlot},
  {Contini}, {Cucciati}, {Garilli}, {Zamorani}, {Adami}, {Bardelli}, {Le Brun},
  {Lemaux}, {Maccagni}, {Pollo}, {Pozzetti}, {Tresse}, {Vergani}, {Zanichelli},
  \& {Zucca}}]{cassata2013}
{Cassata} P. {et~al.}, 2013, \aap, 556, A68

\bibitem[{{Choudhury} \& {Ferrara}(2005)}]{trc2005}
{Choudhury} T.~R., {Ferrara} A., 2005, \mnras, 361, 577

\bibitem[{{Choudhury} \& {Ferrara}(2006)}]{trc2006}
{Choudhury} T.~R., {Ferrara} A., 2006, \mnras, 371, L55

\bibitem[{{Choudhury}, {Srianand} \& {Padmanabhan}(2001){Choudhury},
  {Srianand}, \& {Padmanabhan}}]{tirthanandtp1}
{Choudhury} T.~R., {Srianand} R., {Padmanabhan} T., 2001, \apj, 559, 29

\bibitem[{{Ciardi} {et~al}\mbox{.}(2012){Ciardi}, {Bolton}, {Maselli}, \&
  {Graziani}}]{ciardi}
{Ciardi} B., {Bolton} J.~S., {Maselli} A., {Graziani} L., 2012, \mnras, 423,
  558

\bibitem[{{Eisenstein} \& {Hu}(1999)}]{eisenstein}
{Eisenstein} D.~J., {Hu} W., 1999, \apj, 511, 5

\bibitem[{{Fan}, {Carilli} \& {Keating}(2006){Fan}, {Carilli}, \&
  {Keating}}]{fan}
{Fan} X., {Carilli} C.~L., {Keating} B., 2006, \araa, 44, 415

\bibitem[{{Fanidakis} {et~al}\mbox{.}(2013){Fanidakis}, {Macci{\`o}}, {Baugh},
  {Lacey}, \& {Frenk}}]{fanidakis2013}
{Fanidakis} N., {Macci{\`o}} A.~V., {Baugh} C.~M., {Lacey} C.~G., {Frenk}
  C.~S., 2013, \mnras, 436, 315

\bibitem[{{Fardal}, {Giroux} \& {Shull}(1998){Fardal}, {Giroux}, \&
  {Shull}}]{fardal1998}
{Fardal} M.~A., {Giroux} M.~L., {Shull} J.~M., 1998, \aj, 115, 2206

\bibitem[{{Faucher-Gigu{\`e}re} {et~al}\mbox{.}(2008){Faucher-Gigu{\`e}re},
  {Lidz}, {Hernquist}, \& {Zaldarriaga}}]{faucher2008}
{Faucher-Gigu{\`e}re} C.-A., {Lidz} A., {Hernquist} L., {Zaldarriaga} M., 2008,
  \apj, 688, 85

\bibitem[{{Fukugita} \& {Kawasaki}(1994)}]{fukugita}
{Fukugita} M., {Kawasaki} M., 1994, \mnras, 269, 563

\bibitem[{{Furlanetto}(2009)}]{furlanetto2009}
{Furlanetto} S.~R., 2009, \apj, 703, 702

\bibitem[{{Furlanetto} \& {Oh}(2008)}]{furlanetto2008}
{Furlanetto} S.~R., {Oh} S.~P., 2008, \apj, 682, 14

\bibitem[{{Furlanetto}, {Oh} \& {Briggs}(2006){Furlanetto}, {Oh}, \&
  {Briggs}}]{furlanettorev}
{Furlanetto} S.~R., {Oh} S.~P., {Briggs} F.~H., 2006, \physrep, 433, 181

\bibitem[{{Guimar{\~a}es} {et~al}\mbox{.}(2007){Guimar{\~a}es}, {Petitjean},
  {Rollinde}, {de Carvalho}, {Djorgovski}, {Srianand}, {Aghaee}, \&
  {Castro}}]{guimaraes2007}
{Guimar{\~a}es} R., {Petitjean} P., {Rollinde} E., {de Carvalho} R.~R.,
  {Djorgovski} S.~G., {Srianand} R., {Aghaee} A., {Castro} S., 2007, \mnras,
  377, 657

\bibitem[{{Gunn} \& {Peterson}(1965)}]{gunnpeterson}
{Gunn} J.~E., {Peterson} B.~A., 1965, \apj, 142, 1633

\bibitem[{{Haardt} \& {Madau}(2012)}]{hm12}
{Haardt} F., {Madau} P., 2012, \apj, 746, 125

\bibitem[{{Haiman} \& {Cen}(2002)}]{haimancen2002}
{Haiman} Z., {Cen} R., 2002, \apj, 578, 702

\bibitem[{{Hui} \& {Gnedin}(1997)}]{huignedin}
{Hui} L., {Gnedin} N.~Y., 1997, \mnras, 292, 27

\bibitem[{{Icke}(1979)}]{icke1979}
{Icke} V., 1979, \apj, 234, 615

\bibitem[{{Inoue} {et~al}\mbox{.}(2011){Inoue}, {Kousai}, {Iwata}, {Matsuda},
  {Nakamura}, {Horie}, {Hayashino}, {Tapken}, {Akiyama}, {Noll}, {Yamada},
  {Burgarella}, \& {Nakamura}}]{inoue2011}
{Inoue} A.~K. {et~al.}, 2011, \mnras, 411, 2336

\bibitem[{{Ishida}, {de Souza} \& {Ferrara}(2011){Ishida}, {de Souza}, \&
  {Ferrara}}]{ishida2011}
{Ishida} E.~E.~O., {de Souza} R.~S., {Ferrara} A., 2011, \mnras, 418, 500

\bibitem[{{Jakobsen} {et~al}\mbox{.}(1994){Jakobsen}, {Boksenberg},
  {Deharveng}, {Greenfield}, {Jedrzejewski}, \& {Paresce}}]{jakobsen1994}
{Jakobsen} P., {Boksenberg} A., {Deharveng} J.~M., {Greenfield} P.,
  {Jedrzejewski} R., {Paresce} F., 1994, \nat, 370, 35

\bibitem[{{Jimenez} \& {Haiman}(2006)}]{jimenez2006}
{Jimenez} R., {Haiman} Z., 2006, \nat, 440, 501

\bibitem[{{Kashikawa} {et~al}\mbox{.}(2006){Kashikawa}, {Shimasaku}, {Malkan},
  {Doi}, {Matsuda}, {Ouchi}, {Taniguchi}, {Ly}, {Nagao}, {Iye}, {Motohara},
  {Murayama}, {Murozono}, {Nariai}, {Ohta}, {Okamura}, {Sasaki}, {Shioya}, \&
  {Umemura}}]{kashikawa2006}
{Kashikawa} N. {et~al.}, 2006, \apj, 648, 7

\bibitem[{{Khaire} \& {Srianand}(2013)}]{khaire2013}
{Khaire} V., {Srianand} R., 2013, \mnras, 431, L53

\bibitem[{{Kim} {et~al}\mbox{.}(2009){Kim}, {Stiavelli}, {Trenti}, {Pavlovsky},
  {Djorgovski}, {Scarlata}, {Stern}, {Mahabal}, {Thompson}, {Dickinson},
  {Panagia}, \& {Meylan}}]{kim2009}
{Kim} S. {et~al.}, 2009, \apj, 695, 809

\bibitem[{{Kistler} {et~al}\mbox{.}(2009){Kistler}, {Y{\"u}ksel}, {Beacom},
  {Hopkins}, \& {Wyithe}}]{kistler2009}
{Kistler} M.~D., {Y{\"u}ksel} H., {Beacom} J.~F., {Hopkins} A.~M., {Wyithe}
  J.~S.~B., 2009, \apjl, 705, L104

\bibitem[{{Komatsu} {et~al}\mbox{.}(2011){Komatsu}, {Smith}, {Dunkley},
  {Bennett}, {Gold}, {Hinshaw}, {Jarosik}, {Larson}, {Nolta}, {Page},
  {Spergel}, {Halpern}, {Hill}, {Kogut}, {Limon}, {Meyer}, {Odegard}, {Tucker},
  {Weiland}, {Wollack}, \& {Wright}}]{wmap7}
{Komatsu} E. {et~al.}, 2011, \apjs, 192, 18

\bibitem[{{Larson} {et~al}\mbox{.}(2011){Larson}, {Dunkley}, {Hinshaw},
  {Komatsu}, {Nolta}, {Bennett}, {Gold}, {Halpern}, {Hill}, {Jarosik}, {Kogut},
  {Limon}, {Meyer}, {Odegard}, {Page}, {Smith}, {Spergel}, {Tucker}, {Weiland},
  {Wollack}, \& {Wright}}]{larson}
{Larson} D. {et~al.}, 2011, \apjs, 192, 16

\bibitem[{{Masters} {et~al}\mbox{.}(2012){Masters}, {Capak}, {Salvato},
  {Civano}, {Mobasher}, {Siana}, {Hasinger}, {Impey}, {Nagao}, {Trump},
  {Ikeda}, {Elvis}, \& {Scoville}}]{masters2012}
{Masters} D. {et~al.}, 2012, \apj, 755, 169

\bibitem[{{McDonald} {et~al}\mbox{.}(2001){McDonald}, {Miralda-Escud{\'e}},
  {Rauch}, {Sargent}, {Barlow}, \& {Cen}}]{mcdonald2001}
{McDonald} P., {Miralda-Escud{\'e}} J., {Rauch} M., {Sargent} W.~L.~W.,
  {Barlow} T.~A., {Cen} R., 2001, \apj, 562, 52

\bibitem[{{McQuinn}(2012)}]{mcquinn2012}
{McQuinn} M., 2012, \mnras, 426, 1349

\bibitem[{{Mitra}, {Choudhury} \& {Ferrara}(2011){Mitra}, {Choudhury}, \&
  {Ferrara}}]{mitra2011}
{Mitra} S., {Choudhury} T.~R., {Ferrara} A., 2011, \mnras, 413, 1569

\bibitem[{{Mitra}, {Choudhury} \& {Ferrara}(2012){Mitra}, {Choudhury}, \&
  {Ferrara}}]{mitra2012}
{Mitra} S., {Choudhury} T.~R., {Ferrara} A., 2012, \mnras, 419, 1480

\bibitem[{{Mo}, {van den Bosch} \& {White}(2010){Mo}, {van den Bosch}, \&
  {White}}]{moandwhite}
{Mo} H., {van den Bosch} F.~C., {White} S., 2010, {Galaxy Formation and
  Evolution}

\bibitem[{{Morselli} {et~al}\mbox{.}(2014){Morselli}, {Mignoli}, {Gilli},
  {Vignali}, {Comastri}, {Sani}, {Cappelluti}, {Zamorani}, {Brusa}, {Gallozzi},
  \& {Vanzella}}]{morselli2014}
{Morselli} L. {et~al.}, 2014, arXiv:1406.3961

\bibitem[{{Mortlock} {et~al}\mbox{.}(2011){Mortlock}, {Warren}, {Venemans},
  {Patel}, {Hewett}, {McMahon}, {Simpson}, {Theuns}, {Gonz{\'a}les-Solares},
  {Adamson}, {Dye}, {Hambly}, {Hirst}, {Irwin}, {Kuiper}, {Lawrence}, \&
  {R{\"o}ttgering}}]{mortlock2011}
{Mortlock} D.~J. {et~al.}, 2011, \nat, 474, 616

\bibitem[{{Nakamura} {et~al}\mbox{.}(2011){Nakamura}, {Inoue}, {Hayashino},
  {Horie}, {Kousai}, {Fujii}, \& {Matsuda}}]{nakamura2011}
{Nakamura} E., {Inoue} A.~K., {Hayashino} T., {Horie} M., {Kousai} K., {Fujii}
  T., {Matsuda} Y., 2011, \mnras, 412, 2579

\bibitem[{{Oh} {et~al}\mbox{.}(2001){Oh}, {Nollett}, {Madau}, \&
  {Wasserburg}}]{ohetal2001}
{Oh} S.~P., {Nollett} K.~M., {Madau} P., {Wasserburg} G.~J., 2001, \apjl, 562,
  L1

\bibitem[{{Olive} \& {Skillman}(2004)}]{oliveskillman}
{Olive} K.~A., {Skillman} E.~D., 2004, \apj, 617, 29

\bibitem[{{Ouchi} {et~al}\mbox{.}(2010){Ouchi}, {Shimasaku}, {Furusawa},
  {Saito}, {Yoshida}, {Akiyama}, {Ono}, {Yamada}, {Ota}, {Kashikawa}, {Iye},
  {Kodama}, {Okamura}, {Simpson}, \& {Yoshida}}]{ouchi2010}
{Ouchi} M. {et~al.}, 2010, \apj, 723, 869

\bibitem[{{Planck Collaboration} {et~al}\mbox{.}(2013){Planck Collaboration},
  {Ade}, {Aghanim}, {Armitage-Caplan}, {Arnaud}, {Ashdown}, {Atrio-Barandela},
  {Aumont}, {Baccigalupi}, {Banday}, \& et~al.}]{planck}
{Planck Collaboration} {et~al.}, 2013, arXiv:1303.5076

\bibitem[{{Press} {et~al}\mbox{.}(1992){Press}, {Teukolsky}, {Vetterling}, \&
  {Flannery}}]{nrecipes}
{Press} W.~H., {Teukolsky} S.~A., {Vetterling} W.~T., {Flannery} B.~P., 1992,
  {Numerical recipes in FORTRAN. The art of scientific computing}

\bibitem[{{Pritchard} \& {Loeb}(2010)}]{pritchard2010}
{Pritchard} J., {Loeb} A., 2010, \nat, 468, 772

\bibitem[{{Pritchard}, {Loeb} \& {Wyithe}(2010){Pritchard}, {Loeb}, \&
  {Wyithe}}]{pritchard2010proc}
{Pritchard} J.~R., {Loeb} A., {Wyithe} S., 2010, in Bulletin of the American
  Astronomical Society, Vol.~42, American Astronomical Society Meeting
  Abstracts 215, p. 460.12

\bibitem[{{Raskutti} {et~al}\mbox{.}(2012){Raskutti}, {Bolton}, {Wyithe}, \&
  {Becker}}]{raskutti2012}
{Raskutti} S., {Bolton} J.~S., {Wyithe} J.~S.~B., {Becker} G.~D., 2012, \mnras,
  421, 1969

\bibitem[{{Reimers} {et~al}\mbox{.}(2005){Reimers}, {Fechner}, {Hagen},
  {Jakobsen}, {Tytler}, \& {Kirkman}}]{reimers2005}
{Reimers} D., {Fechner} C., {Hagen} H.-J., {Jakobsen} P., {Tytler} D.,
  {Kirkman} D., 2005, \aap, 442, 63

\bibitem[{{Ricotti}, {Gnedin} \& {Shull}(2000){Ricotti}, {Gnedin}, \&
  {Shull}}]{ricotti2000a}
{Ricotti} M., {Gnedin} N.~Y., {Shull} J.~M., 2000, \apj, 534, 41

\bibitem[{{Robertson} \& {Ellis}(2012)}]{robertsonellis2012}
{Robertson} B.~E., {Ellis} R.~S., 2012, \apj, 744, 95

\bibitem[{{Rollinde} {et~al}\mbox{.}(2005){Rollinde}, {Srianand}, {Theuns},
  {Petitjean}, \& {Chand}}]{rollinde2005}
{Rollinde} E., {Srianand} R., {Theuns} T., {Petitjean} P., {Chand} H., 2005,
  \mnras, 361, 1015

\bibitem[{{Schaye} {et~al}\mbox{.}(2000){Schaye}, {Theuns}, {Rauch},
  {Efstathiou}, \& {Sargent}}]{schaye2000}
{Schaye} J., {Theuns} T., {Rauch} M., {Efstathiou} G., {Sargent} W.~L.~W.,
  2000, \mnras, 318, 817

\bibitem[{{Seljak}, {Slosar} \& {McDonald}(2006){Seljak}, {Slosar}, \&
  {McDonald}}]{seljak2006}
{Seljak} U., {Slosar} A., {McDonald} P., 2006, \jcap, 10, 14

\bibitem[{{Shull} {et~al}\mbox{.}(2010){Shull}, {France}, {Danforth}, {Smith},
  \& {Tumlinson}}]{shull2010}
{Shull} J.~M., {France} K., {Danforth} C.~W., {Smith} B., {Tumlinson} J., 2010,
  \apj, 722, 1312

\bibitem[{{Springel}(2005)}]{gadget2}
{Springel} V., 2005, \mnras, 364, 1105

\bibitem[{{Stark} {et~al}\mbox{.}(2007){Stark}, {Ellis}, {Richard}, {Kneib},
  {Smith}, \& {Santos}}]{stark2007}
{Stark} D.~P., {Ellis} R.~S., {Richard} J., {Kneib} J.-P., {Smith} G.~P.,
  {Santos} M.~R., 2007, \apj, 663, 10

\bibitem[{Syphers \& Shull(2014)}]{syphers2014}
Syphers D., Shull J.~M., 2014, The Astrophysical Journal, 784, 42

\bibitem[{{Theuns} {et~al}\mbox{.}(2002){Theuns}, {Schaye}, {Zaroubi}, {Kim},
  {Tzanavaris}, \& {Carswell}}]{theuns2002}
{Theuns} T., {Schaye} J., {Zaroubi} S., {Kim} T.-S., {Tzanavaris} P.,
  {Carswell} B., 2002, \apjl, 567, L103

\bibitem[{{Theuns} \& {Zaroubi}(2000)}]{theuns2000}
{Theuns} T., {Zaroubi} S., 2000, \mnras, 317, 989

\bibitem[{{Tornatore}, {Ferrara} \& {Schneider}(2007){Tornatore}, {Ferrara}, \&
  {Schneider}}]{tornatore2007}
{Tornatore} L., {Ferrara} A., {Schneider} R., 2007, \mnras, 382, 945

\bibitem[{{Totani} {et~al}\mbox{.}(2006){Totani}, {Kawai}, {Kosugi}, {Aoki},
  {Yamada}, {Iye}, {Ohta}, \& {Hattori}}]{totani2006}
{Totani} T., {Kawai} N., {Kosugi} G., {Aoki} K., {Yamada} T., {Iye} M., {Ohta}
  K., {Hattori} T., 2006, \pasj, 58, 485

\bibitem[{{Venkatesan}, {Tumlinson} \& {Shull}(2003){Venkatesan}, {Tumlinson},
  \& {Shull}}]{venkatesan2003}
{Venkatesan} A., {Tumlinson} J., {Shull} J.~M., 2003, \apj, 584, 621

\bibitem[{{Viel}, {Haehnelt} \& {Lewis}(2006){Viel}, {Haehnelt}, \&
  {Lewis}}]{viel2006}
{Viel} M., {Haehnelt} M.~G., {Lewis} A., 2006, \mnras, 370, L51

\bibitem[{{Walter} {et~al}\mbox{.}(2003){Walter}, {Bertoldi}, {Carilli}, {Cox},
  {Lo}, {Neri}, {Fan}, {Omont}, {Strauss}, \& {Menten}}]{walter2003}
{Walter} F. {et~al.}, 2003, \nat, 424, 406

\bibitem[{{Willott} {et~al}\mbox{.}(2007){Willott}, {Delorme}, {Omont},
  {Bergeron}, {Delfosse}, {Forveille}, {Albert}, {Reyl{\'e}}, {Hill},
  {Gully-Santiago}, {Vinten}, {Crampton}, {Hutchings}, {Schade}, {Simard},
  {Sawicki}, {Beelen}, \& {Cox}}]{willott2007}
{Willott} C.~J. {et~al.}, 2007, \aj, 134, 2435

\bibitem[{{Willott} {et~al}\mbox{.}(2005){Willott}, {Percival}, {McLure},
  {Crampton}, {Hutchings}, {Jarvis}, {Sawicki}, \& {Simard}}]{willott2005}
{Willott} C.~J., {Percival} W.~J., {McLure} R.~J., {Crampton} D., {Hutchings}
  J.~B., {Jarvis} M.~J., {Sawicki} M., {Simard} L., 2005, \apj, 626, 657

\bibitem[{{Worseck} {et~al}\mbox{.}(2011){Worseck}, {Prochaska}, {McQuinn},
  {Dall'Aglio}, {Fechner}, {Hennawi}, {Reimers}, {Richter}, \&
  {Wisotzki}}]{worseck2011}
{Worseck} G. {et~al.}, 2011, \apjl, 733, L24

\bibitem[{{Wyithe} \& {Bolton}(2011)}]{wyithe2011}
{Wyithe} J.~S.~B., {Bolton} J.~S., 2011, \mnras, 412, 1926

\bibitem[{{Wyithe} \& {Loeb}(2003)}]{wyithe2003}
{Wyithe} J.~S.~B., {Loeb} A., 2003, \apj, 586, 693

\bibitem[{{Zheng} {et~al}\mbox{.}(2004){Zheng}, {Kriss}, {Deharveng}, {Dixon},
  {Kruk}, {Shull}, {Giroux}, {Morton}, {Williger}, {Friedman}, \&
  {Moos}}]{zheng2004}
{Zheng} W. {et~al.}, 2004, \apj, 605, 631

\end{thebibliography}

\appendix

\section{Calculational details}
\label{appendixa}
In this appendix, we present the details of the calculations which were described briefly in Section \ref{sec:nummodel}.
The initial conditions are described by photoionization equilibrium with the background, this system of equations is
given by:
\begin{eqnarray}
n_{\rm{HI}} \Gamma^{\rm bg}_{\rm{HI}} &=& n_{\rm{HII}} n_e \alpha_{\rm{HII}}  \nonumber
\\
 n_{\rm{HeI}} \Gamma^{\rm bg}_{\rm{HeI}} + n_{\rm{HeIII}} n_e \alpha_{\rm{HeIII}} &=&
n_{\rm{HeII}} \Gamma^{\rm bg}_{\rm{HeII}} + n_{\rm{HeII}} n_e \alpha_{\rm{HeII}}
\nonumber \\
 n_{\rm{HeII}} \Gamma^{\rm bg}_{\rm{HeII}} &=& n_{\rm{HeIII}} n_e \alpha_{\rm{HeIII}}  
\label{eqhm}
\end{eqnarray}
with the boundary conditions that: $n_{\rm HI} + n_{\rm HII} = n_{\rm H}$,
$n_{\rm HeI} + n_{\rm HeII} + n_{\rm HeIII} = n_{\rm He}$, and $n_{\rm HII} +
n_{\rm HeII} + 2 n_{\rm HeIII} = n_{e}$. Here, $\Gamma^{\rm bg}_{\rm x}$ represents the photoionization rate of species `x' from the
background ionizing radiation assumed \citep[we use][]{hm12}, the $\alpha$'s are
the radiative recombination rate coefficients, and the $n$'s represent the
(proper) number densities.

The
background photoionization rates are given by (in ${\rm{s}}^{-1}$):
\begin{eqnarray}
 \Gamma^{\rm bg}_{\rm HI} = 2.30 \times 10^{-13} ; \ \Gamma^{\rm bg}_{\rm HeI} = 1.54 \times
10^{-13}; \nonumber \\ 
\Gamma^{\rm bg}_{\rm HeII} = 4.42 \times 10^{-19} .
 \label{hmgamma6}
\end{eqnarray} 

The temperatures are assigned to each pixel by the equation of state:
\begin{equation}
 T(x,z) = T_0(z) [1 + \delta(x)]^{\gamma - 1}
\label{eos}
\end{equation} 
where, $T_0$ is the normalization temperature, $\delta(x)$ is the overdensity at
the pixel and $\gamma$ is the slope of the equation of state. 

 Our numerical procedure now involves solving the system of
four differential equations for the temperature evolution and hydrogen and
helium ion densities evolution:
\begin{eqnarray}
 \frac{d n_{\rm{HII}}}{dt} &=& n_{\rm{HI}} \Gamma_{\rm{HI}} - n_{\rm{HII}} n_e 
\alpha_{\rm{HII}} - 3H(t)n_{\rm HII}\nonumber \\
 \frac{d n_{\rm{HeII}}}{dt} &=& n_{\rm{HeI}} \Gamma_{\rm{HeI}} + n_{\rm{HeIII}}
n_e \alpha_{\rm{HeIII}}  \nonumber\\
\qquad &-& n_{\rm{HeII}} \Gamma_{\rm{HeII}} -n_{\rm{HeII}} n_e
\alpha_{\rm{HeII}} - 3H(t)n_{\rm HeII} \nonumber \\
 \frac{d n_{\rm{HeIII}}}{dt} &=& n_{\rm{HeII}} \Gamma_{\rm{HeII}} -
n_{\rm{HeIII}} n_e \alpha_{\rm{HeIII}} - 3H(t)n_{\rm HeIII} \nonumber \\
 \frac{dT}{dt} &=& \frac{2}{3 k_B n_{tot}} [H_{tot}(n_i) -
C(n_i,T)] \nonumber \\
               &&\qquad - 2 H(t) T - \frac{T}{n_{tot}}\frac{d n_{tot}}{dt}
 \label{evoleqns}
\end{eqnarray}
In the above equations, $\Gamma_{\rm x}$ represents the photoionization rates of
species `x' (contributed both by the quasar as well as the background in the
near-zone, and by the background alone, for the far zone). The adiabatic index
is $5/3$,  and
$H_{tot}(n_i)$ and $C(n_i,T)$ represent the total photoheating rate per unit
volume, and radiative cooling function respectively. The Hubble parameter is $H(t)$, and  $n_{tot} = n_{\rm H} + n_{\rm He} + n_e$ is the total number
density of particles of different species. The term $- 2 H(t) T$ in the
temperature evolution equation represents the contribution of the expansion of
the universe to the adiabatic cooling of the gas. We ignore the contribution from $- 3H(t)n$ in the evolution of the species densities, since the ionization time scales under consideration are much smaller than $H^{-1}(t)$.
The last term $-T (d n_{tot}/dt) /n_{tot}$ represents the correction due to species evolution. This correction is only
about 1 part in $10^3$ at the highest temperatures, but the effect is
expected to be important in the initial stages of evolution.

The photoionization rates from the quasar at a distance $R$ are given by:

\begin{eqnarray}
 \Gamma_{\rm HI}^{QSO} (R) &=& \int_{\nu_{\rm HI}}^{\infty} \frac{L_{\nu}}{4 \pi
R^2 h \nu} \sigma_{\rm HI} (\nu) \exp(-\tau_{\rm HI}) \ d \nu ; \nonumber \\
 \Gamma_{\rm HeI}^{QSO} (R) &=& \int_{\nu_{\rm HeI}}^{\infty} \frac{L_{\nu}}{4
\pi R^2 h \nu} \sigma_{\rm HeI} (\nu) \exp(-\tau_{\rm HeI}) \ d \nu\nonumber \\
 \Gamma_{\rm HeII}^{QSO} (R) &=& \int_{\nu_{\rm HeII}}^{\infty} \frac{L_{\nu}}{4
\pi R^2 h \nu} \sigma_{\rm HeII} (\nu) \exp(-\tau_{\rm HeII}) \ d \nu
\label{qsophotoion}
\end{eqnarray} 
where $L_{\nu} = L_{\rm HI} (\nu/\nu_{\rm HI})^{-\alpha_s}$. The $\sigma(\nu)$'s denote the photoionization cross-sections for \HI, \HeI\ and
\HeII\ respectively and the $\tau$'s are the  corresponding optical depths,
 calculated as
 \begin{eqnarray}
\tau_{\rm x}(R) &=& \sum_{i = 1}^{n(R)}[n_{\rm HI} (i) \sigma_{\rm HI}  (\nu_{\rm x}) + n_{\rm HeI} (i) \sigma_{\rm HeI} (\nu_{\rm x}) \nonumber\\
&+& n_{\rm HeII} (i) \sigma_{\rm HeII} (\nu_{\rm x})] l                                                                                                                                                                                           \end{eqnarray} 
where, $l$ is the pixel size and $\nu_{\rm x}$ is the ionization edge of species $\rm{x} =$ \HI, \HeI\ or \HeII.  For simplicity of computation, we only consider the optical depth at the ionization edge of the relevant species in the photoionization rate. The sum is over all the pixels up to the $n(R)$th pixel which is at the distance $R$ from the quasar. The total photoionization rate  is obtained by adding the contributions from the quasar [\eq{qsophotoion}] and the metagalactic background [\eq{hmgamma6}].
 
Recombination rates are as given in \citet{fukugita}, \citet{abel} and
\citet{moandwhite} for \HII, \HeII\ (including dielectronic recombination) and
\HeIII. We use case A recombination
coefficients here as they have been found to be the appropriate choice for
comparison with hydrodynamical simulations of quasar near-zones \citep{bolton07}. The details are as follows:

\begin{enumerate}
\item Case A recombination coefficients (in $\rm{cm}^3 \rm{s}^{-1}$):

(a) $\alpha_{\rm HII} = 6.28 \times 10^{-11} T^{-0.5} (T/1000)^{-0.2} (1 +
(10^{-6} T)^{0.7})^{-1}$

(b) $\alpha_{\rm HeII} = 1.5 \times 10^{-10} T^{-0.6353}$

(c) $\alpha_{\rm HeIII} = 3.3 \times 10^{-10} T^{-0.5} (T/1000)^{-0.2} (1 + (2.5
\times 10^{-7} T)^{0.7})^{-1} $

\item Dielectronic recombination coefficient for helium (in $\rm{cm}^3 \rm{s}^{-1}$):

(a)  $\alpha_{\rm HeII}^d = 1.93 \times 10^{-3} T^{-1.5} \exp (-470000/T) (1 +
0.3 \exp(-94000/T))$
\end{enumerate}

To analyze the photo-heating, we use the background heating rates as given in
\citet{hm12} at redshift $\sim 6$ (in ergs s$^{-1}$):
\begin{eqnarray}
 && E^{\rm bg}_{\rm HI} = 1.5824 \times 10^{-24} ; \ E^{\rm bg}_{\rm HeI} = 1.792 \times 10^{-24};
\nonumber \\
 &&  E^{\rm bg}_{\rm HeII} = 4.304 \times 10^{-29} \,.
\end{eqnarray} 
We add to the above background heating rates, the additional heating rate due to
the quasar with the previously mentioned luminosity and spectral index, given
by:
\begin{eqnarray}
 E_{\rm HI}^{QSO} (R) &=& \int_{\nu_{\rm HI}}^{\infty} \frac{L_{\nu} h (\nu -
\nu_{\rm HI})}{4 \pi R^2 h \nu} \sigma_{\rm HI} (\nu) \exp(-\tau_{\rm HI}) \ d \nu ;
\nonumber \\
 E_{\rm HeI}^{QSO} (R) &=& \int_{\nu_{\rm HeI}}^{\infty} \frac{L_{\nu} h (\nu -
\nu_{\rm HeI}) }{4 \pi R^2 h \nu} \sigma_{\rm HeI} (\nu) \exp(-\tau_{\rm HeI})\ d \nu
\nonumber \\
 E_{\rm HeII}^{QSO} (R) &=& \int_{\nu_{\rm HeII}}^{\infty} \frac{L_{\nu} h (\nu
- \nu_{\rm HeII})}{4 \pi R^2 h \nu} \sigma_{\rm HeII} (\nu) \exp(-\tau_{\rm HeII}) \ d \nu
\nonumber\\
\end{eqnarray} 
At any distance $R$ from the quasar, the total photoheating rate per unit
volume, $H_{tot} (R)$, is given by $H_{tot} (R) = \sum_{i} n_i [E_i^{\rm bg} +
E_i^{QSO}(R)]$ where the sum is over $i =$ \HI, \HeI\ and \HeII.

The cooling function consists of contributions from (a) bremsstrahlung and (b)
recombination. We use the corresponding expressions as given by \citet{fukugita},
\citet{abel} and \citet{moandwhite} for \HII, \HeII\ and \HeIII, including a
contribution from the dielectronic recombination of \HeII. Collisional ionization
and its associated cooling are ignored since, for the range of temperatures and
densities considered here, their magnitudes are negligible as compared to the photoionization
and the cooling rates by recombination and bremsstrahlung respectively, which we
have considered here. The details are:

\begin{enumerate}
\item Recombination cooling rates (in erg $\rm{cm}^{-3} \rm{s}^{-1}$):
 
(a) $\Lambda_{\rm HII} = 2.82 \times 10^{-26} T^{0.3} ( 1 + 3.54 \times 10^{-6}
T)^{-1} n_{\rm HII} n_e$

(b) $\Lambda_{\rm HeII} = 1.55 \times 10^{-26} T^{0.3647} n_{\rm HeII} n_e $

(c) $\Lambda_{\rm HeIII} = 1.49 \times 10^{-25} T^{0.3} (1 + 0.885 \times
10^{-6} T)^{-1} n_{\rm HeIII} n_e$

\item Dielectronic recombination cooling rate for helium (in erg $\rm{cm}^{-3} \rm{s}^{-1}$):

(a) $\Lambda_{\rm HeII}^d = 1.24 \times 10^{-13} T^{-1.5} \exp (-470000/T) (1 +
0.3 \exp(-94000/T)) n_{\rm HeII} n_e$

\item Bremsstrahlung (in erg $\rm{cm}^{-3} \rm{s}^{-1}$):

$\Lambda_b = 1.43 \times 10^{-27} T^{0.5} \ g_{\rm ff} n_e (n_{\rm HII} + n_{\rm
HeII} + 4 n_{\rm HeIII})$
where, the Gaunt factor $g_{\rm ff}$ is given by $g_{\rm ff} = 1.1 + 0.34
\exp(-(5.5. - \rm{log}_{10} T)^2/3)$.

\end{enumerate}

\subsection{Description of the code}
\label{sec:codedesc}

\begin{figure*}
 \begin{center}
  \includegraphics[scale=0.78]{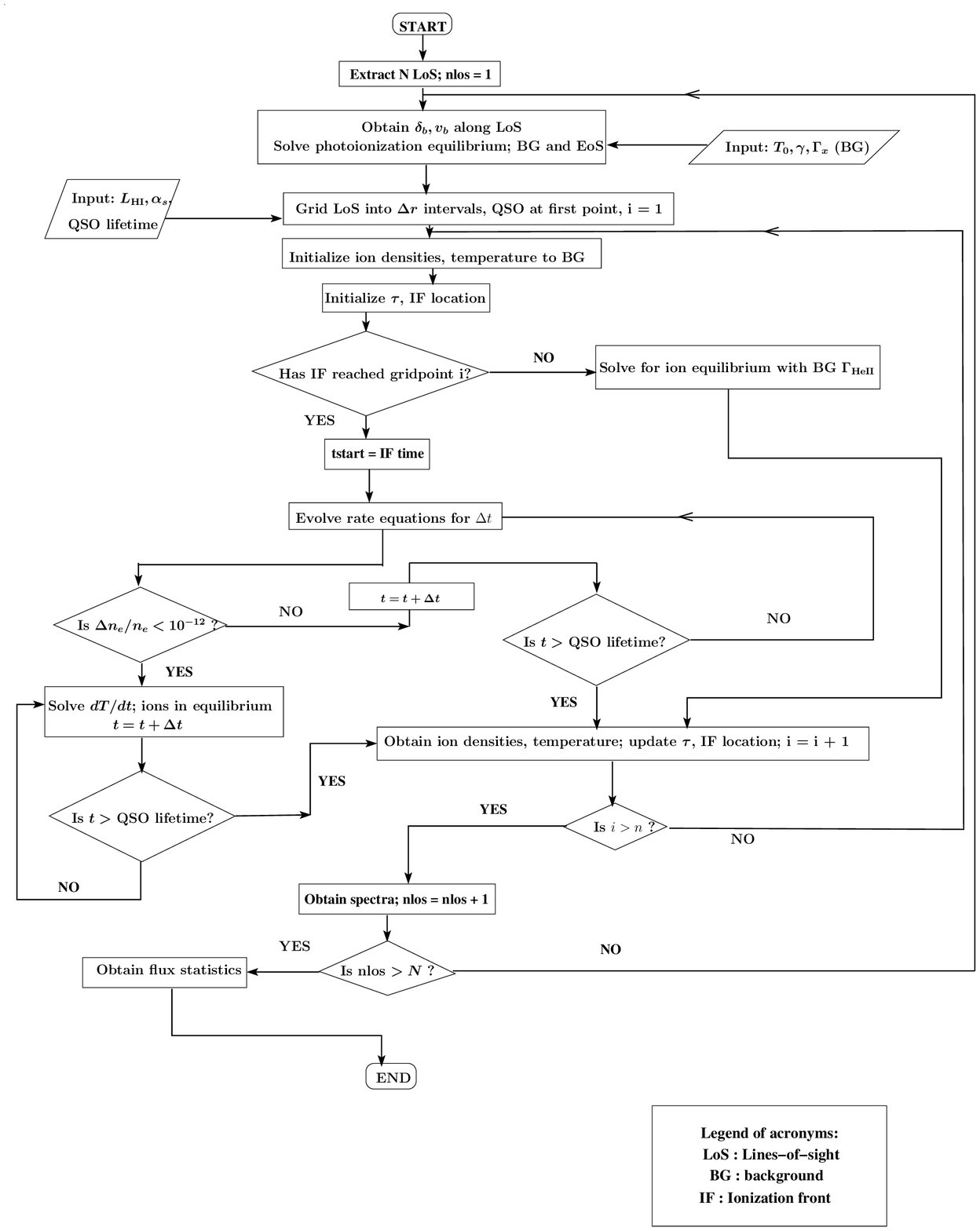}
  \caption{Flowchart describing the numerical scheme.}
  \label{fig:flowchart}
 \end{center}
\end{figure*}

The algorithmic procedure is as outlined in \fig{fig:flowchart}. First, a number $N$ lines
of sight  are extracted randomly in our simulation box at redshift 6. For each
line of sight, the density and velocity fields, $\delta_b$, and $v_b$ of the baryonic particles are obtained.
The equilibrium ion number densities and temperature are found under the
assumption of photoionization with the metagalactic background and equation of state, by solving \eq{eqhm} using
the Newton-Raphson technique with the routine NEWT in Numerical
Recipes \citep{nrecipes}.  The inputs to the code at this stage are $T_0$, $\gamma$ and the background photoionization rates.

Next, the line-of-sight is gridded into $n$ equispaced intervals with the length of each interval being equal to the average pixel size in the simulation, and the quasar is placed at the first gridpoint.
The inputs are the luminosity of the quasar at the Lyman-edge, $L_{\rm HI}$, the
spectral index $\alpha_s$ and the quasar lifetime $t_Q$. The start time of evolution of the thermal and ionization state of a
gridpoint interval is decided by the time at which the \HeII\ ionization front
reaches that gridpoint, which is calculated from \eq{front} using the known distance to the gridpoint\footnote{Strictly speaking, one should evolve the gridpoint even if the \HeII\ front has not reached it, to account for the Hubble expansion. However, we do not do this since the time scales under present consideration are much shorter than $H^{-1}(t)$.}. The initial conditions are the equilibrium species
fractions and temperatures found previously. The four rate equations in
\eq{evoleqns} are now solved using a FORTRAN90 code based on the ODEINT routine of
the Numerical Recipes \citep{nrecipes}. The ion densities and temperatures at
each gridpoint interval are evolved with a time-step $\Delta t$, which is
dynamic in nature, being inversely proportional to the rate of ionizing photons
at the distance of the gridpoint; a typical value being $\Delta t\sim 10^6$ s. We follow the approach of \citet{bolton07} in that when the relative change in the electron
number density falls below $10^{-12}$, the ion fractions are solved for
assuming 
photoionization equilibrium and a larger time-step is considered. 

In case the \HeII\ ionization front has not yet reached a particular gridpoint
within the quasar lifetime, the temperatures and ion densities are solved for
assuming photoionization equilibrium with no contribution from the quasar to
$\Gamma_{\rm HeII}$ and $E_{\rm HeII}$. Hence, those gridpoints
located beyond the \HeII\ front do not ``see'' the quasar as far as
photoionization of \HeII\ and the resulting gas heating are concerned. In this way, 
the location of the \HeII\ ionization front at the end of the quasar lifetime is
also known.

The final values of number densities of different species and the temperature,
for each gridpoint, are then used to update the optical depth values at the ionization edges,
and the location of the \HeII\ ionization front. Once these values are passed to the next
gridpoint, the process is repeated until the end of the
line-of-sight is reached. The temperature and neutral hydrogen
density at each pixel are used to generate the simulated spectrum along that
line-of-sight, by defining the redshift grid as described in the previous
section. Note that in this procedure, the optical depth value contributes to the determination of the location of the ionization front, which determines the start time of the next gridpoint and its consequent evolution, which in turn contributes to the optical depth for the further gridpoints under consideration. Hence, if the (integrated) optical depth effect becomes large enough so that the front is ``stopped'', the subsequent gridpoints do not ``see'' the ionization and heating photons from the quasar, and are ionized and heated by the background alone. 

Finally, the combined set of all the
gridpoints at each line of sight, and the number of lines of sight extracted in
the simulation box are used to obtain the flux statistics. In our simulations, we do not use the realistic quasar continuum to generate spectra. Hence, all the artificial effects coming from the issues related to continuum fitting will not be present in our analysis.

\end{document}